# Semi-annual, annual and Universal Time variations in the magnetosphere and in geomagnetic activity: 4. Polar Cap motions and origins of the Universal Time effect

Mike Lockwood[1], Carl Haines[1], Luke A. Barnard[1], Mathew J. Owens[1], Chris J. Scott[1], Aude Chambodut[2], and Kathryn A. McWilliams[3]

[1] *Department of Meteorology, University of Reading, Reading, UK*

[2] *École et Observatoire des Sciences de la Terre, Université de Strasbourg and CNRS, France*

[3] *Institute of Space and Atmospheric Studies, University of Saskatchewan, Saskatoon, Saskatchewan, Canada.*

**Abstract.** We use the *am*, *an, as* and the $a\sigma$ geomagnetic indices to explore a previously overlooked factor in magnetospheric electrodynamics, namely the inductive effect of diurnal motions of the Earth's magnetic poles toward and away from the Sun caused by Earth's rotation. Because the offset of the (eccentric dipole) geomagnetic pole from the rotational axis is roughly twice as large in the southern hemisphere compared to the northern, the effects there are predicted to be roughly twice the amplitude of those in the northern hemisphere. Hemispheric differences have previously been discussed in terms of polar ionospheric conductivities generated by solar photoionization, effects which we allow for by looking at the dipole tilt effect on the time-of-year variations of the indices. The electric field induced in a geocentric frame is shown to also be a significant factor and gives a modulation of the voltage applied by the solar wind flow in the southern hemisphere that is typically a ±30% diurnal modulation for disturbed intervals rising to ±76% in quiet times. For the northern hemisphere these are 15% and 38% modulations. Motion towards/away from the Sun reduces/enhances the directly-driven ionospheric voltages and reduces/enhances the magnetic energy stored in the tail and we estimate that approximately 10% of the effect appears in directly driven ionospheric voltages and 90% in changes of the rate of energy storage or release in the near-Earth tail. The hemispheric asymmetry in the geomagnetic pole offsets from the rotational axis is shown to be the dominant factor in driving Universal Time (*UT*) variations and hemispheric differences in geomagnetic activity. Combined with the effect of solar wind dynamic pressure and dipole tilt on the pressure balance in the near-Earth tail, the effect provides an excellent explanation of how the observed Russell-McPherron pattern with time-of-year $F$ and $UT$ in the driving power input into the magnetosphere is converted into the equinoctial $F$ - $UT$ pattern in average geomagnetic activity (after correction is made for dipole tilt effects on ionospheric conductivity), added to a pronounced $UT$ variation with minimum at 02-10$UT$. In addition, we show that the predicted and observed $UT$ variations in average geomagnetic activity has implications for the occurrence of the largest events that also show the nett $UT$ variation.

## 1. Introduction

The first well-informed description of a Universal Time ($UT$) variation in global geomagnetic activity, that we know of, was by *Bartels* (1925, 1928). Bartels postulated that it was linked to the angle of tilt $\psi$ of Earth's magnetic axis relative to the sunward direction ($X$) from studying the "U index" which commenced in 1835 and was continued until the 1930s: until 1871 it was based on declination readings from two observatories, after which it was based on seven stations (*Russell and McPherron,* 1973; *Nevanlinna*, 2004). The U index is equivalent to the magnitude of the difference between successive daily averages of the modern Dst index. In their seminal book *Chapman and Bartels* (1940) commented (section XI.20, page 391), "since the local time of Batavia and Potsdam differ by 5-6 hours, the identity of the hours of maximum or minimum $U$ suggests the existence of a 'Universal Time' variation of $U$. Such a variation might depend, for example, on the varying angle between the Earth's magnetic axis and the line connecting the Sun and the Earth". This idea, now usually referred to as "dipole tilt effects" or the "equinoctial hypothesis", was employed by *Waldo-Lewis and McIntosh* (1953), *McIntosh* (1959) and many authors since (e.g., *Aoki*, 1977). The characteristic equinoctial pattern of variation with time-of-year and $UT$ that this generates is also sometimes said to be caused by the "McIntosh Effect" (e.g., *Berthelier*, 1990). The dipole tilt angle $\psi$ varies with $UT$ because of the rotation of the Earth and the offset of the geomagnetic dipole axis, $\vec{M}$, relative to Earth's rotational axis $\vec{\Omega}$. The tilt angle $\psi$ also varies with time-of-year because of Earth's motion around the Sun (the rotation axis $\vec{\Omega}$ being fixed in the inertial frame but the direction toward the Sun rotating through 360° every year in that frame). Hence the equinoctial hypothesis links a $UT$ variation with annual and semi-annual variations in geomagnetic activity by the precessions of the $\vec{M}$ and $\vec{\Omega}$ axes.

For a geocentric, symmetric dipole, most effects of the dipole tilt vary in amplitude with $|\psi|$ and if this fully applies, effects in one hemisphere are equal and opposite to those in the other hemisphere and the net global effect is zero when averaged over intervals of a whole number of years. However, the geomagnetic field is not a symmetric, Earth-centred dipole (e.g., *Koochak and Fraser-Smith*, 2017) and this will cause *UT* variations even in global data and even when averaged over many years. The third mechanism discussed in relation to the semiannual variation is the "axial effect" (see review in Paper 1, *Lockwood et al.*, 2020a) which depends on the variation of the heliographic latitude of Earth over the year because the ecliptic is inclined at about 7° with respect to the solar equator. This effect alters the probability of Earth intersecting faster solar wind and so can introduces a time-of-year variation but no *UT* effect because it does not invoke variations caused by the orientation of Earth's magnetic field.

As described in Papers 1 and 2 of this series (*Lockwood et al.,* 2020a; b), the semi-annual variation is well explained by the "Russell-McPherron" (R-M) effect (*Russell and McPherron*, 1973) which is due to the effect of the orientation of the interplanetary magnetic field (IMF) on magnetic reconnection in the dayside magnetopause and hence on solar-wind magnetosphere coupling. However, this predicts a pattern of response with fraction of a calendar year ($F$) and *UT* that is very different from the equinoctial pattern that is seen in geomagnetic activity using the best indices that have responses to solar wind forcing that do not vary with either $F$ or *UT*. Several studies have tried to explain the observed equinoctial pattern by amending the R-M theory to include a dipole tilt effect in solar-wind magnetosphere coupling (see review in Paper 1). However, *Finch et al.* (2008) showed that the equinoctial pattern is not found in data from dayside auroral and polar cap magnetometer stations responding to directly-driven currents, and only in data from nightside stations responding to the substorm current wedge. The conclusion is supported by the work *Chambodut et al.* (2013) who showed that in the mid-latitude $a\sigma$ indices, the equinoctial pattern is strongest in the midnight sector and weakest in the noon sector. This shows that the equinoctial pattern is an internal response of the magnetosphere-ionosphere system and not present in solar-wind/magnetosphere coupling (*Lockwood*, 2013). There has been much debate as to whether the R-M and equinoctial effects are separate phenomena (e.g., *Berthelier*, 1990; *Russell and Scurry*, 1990; *de la Sayette and Berthelier*, 1976); however, as reviewed in Papers 1 and 2, the fact that the equinoctial pattern splits into a March and a September peak when the data are sorted by the polarity of the prevailing *Y*-component of the IMF shows that the R-M effect is at the heart of the equinoctial effect (*Berthelier*, 1976; *Nakai*, 1990; *Zhao and Zong*, 2012; *Lockwood et al.*, 2016), it being unique in predicting this division. As reviewed in Paper 1, there have been a large number of theories proposed, but we have not yet developed an understanding of how the characteristic Russell-McPherron *F-UT* pattern in solar wind-magnetosphere coupling evolves into an equinoctial pattern in geomagnetic response. Lastly, it must be remembered that, as demonstrated by *Lockwood et al.* (2016), any activity indices (such as the *Dst* geomagnetic index) that depend on the prior solar wind conditions integrated over timescales longer than about 12 hrs will necessarily show an "axial" pattern (with no clear *UT* dependence) rather than either an R-M or equinoctial pattern.

On averaging over a full year, both the R-M and equinoctial patterns predict that there would be no residual *UT* variation if there is symmetry between the two hemispheres in terms of geomagnetic field and seasonal variations in ionospheric conductivities. However, analysis of geomagnetic data strongly suggests that this is not the case with reports of a persistent minimum at about 3-9 hr *UT*. This was first noted in the Auroral Electrojet indices *AE* and *AL* (*Davis and Siguira*, 1966; *Allen and Kroehl*, 1975, *Basu*, 1975, *Aoki*, 1977) and has been reported many times since (*Hajkowicz*, 1992, 1998; *Ahn et al.*, 2000; *Ahn and Moon*, 2003). However, these are northern-hemisphere indices based on a ring of observing stations around the northern-hemisphere auroral oval and the main limitation of these studies is that without a southern-hemisphere equivalent, variations could be seasonal effects that are cancelled on a global scale by anti-phase seasonal effects in the southern hemisphere. There have been attempts to construct southern hemisphere *AE* indices (*Maclennan et al.*, 1991; *Weygand et al*., 2014) but large parts of the southern auroral oval are over sea or ocean, giving large gaps in longitudinal coverage and so detection of *UT* variations is particularly limited. Any differences in the longitudinal spacing of the stations could introduce a spurious *UT* variation, as could longitudinal variations in the difference in latitude between the average auroral oval location and the stations. Initially there were just 5 stations in the northern hemisphere *AE* ring, and longitudinal coverage was indeed a concern but his was soon increased to 12. As will be discussed below, we now know that the *UT* variation is still present in the equivalent *SME* and *SML* SuperMAG indices (*Newell and Gjerloev,* 2011) derived from of order 100 stations in the northern hemisphere (*Singh et al*., 2013; *Wang et al.*, 2014). Hence, like *Singh* *et al.* (2013), we eliminate the positioning of the *AE* stations as the cause of the *UT* variation; however, the fact that these observing networks are in just one hemisphere remains a relevant factor.

A commonly-used planetary index is $kp$ (equivalent to $ap$) but this is unsuitable for studying *UT* variations as the data are mapped back via conversion tables to the observations made at one station (Niemegk) and *Lockwood et al.* (2019a) have shown this gives the $kp$ and $ap$ indices a large *F-UT* network response variation that makes it unsuitable for detecting *UT* variations. On the other hand, *Lockwood et al.* (2019a) use a model of the response of each station to show that the $am$ index (*Mayaud*, 1967, 1980) has an extremely flat *F-UT* response, especially at higher activity levels. This means that studies that reported a persistent *UT* variation in the $am$ index (*Berthelier*, 1976; *Russell*, 1989; *de La Sayette and Berthelier*, 1996; *Cliver et al.*, 2000) are particularly significant. There are other global *UT* variations in the magnetosphere that have been remotely sensed. For example, *Morioka et al.* (2013) have shown that Auroral Kilometric Radiation (AKR) data has a *UT* variation in frequency and amplitude that is the same in both hemispheres and that this is not related to the visibility of the magnetosphere for the observing GEOTAIL spacecraft that was outside the magnetosphere. The authors infer it is generated by the effect on the auroral acceleration regions of the bending of the tail with the dipole tilt although the precise mechanism remains unclear. In addition, *Luan et al.* (2016) have studied the *UT* variation and hemispheric asymmetry in auroral power deposition using observations by the TIMED (Thermosphere Ionosphere Mesosphere Energetics and Dynamics) satellite.

This is the fourth in a series of papers investigating semi-annual, annual, and *UT* variations in the magnetosphere in which we study *UT* variations in the magnetosphere making use of the global $am$ and $a\sigma$ geomagnetic indices. The $a\sigma$ indices are generated in almost the same way as $am$, but employ only data from one of four 6-hour sectors of Magnetic Local Time (MLT) around dawn, noon, dusk and midnight (*Chambodut et al.*, 2013). We here also study for the first time, the hemispheric sub-indices that are averaged together to generate these global $a\sigma$ indices: this mean that in addition to northern and southern hemisphere indices $an$ and $as$ (where $am = (an + as)/2$), we also employ: $a\sigma_N(dawn)$, $a\sigma_S(dawn)$ and $a\sigma(dawn) = \{a\sigma_N(dawn) + a\sigma_S(dawn)\}/2$); $a\sigma_N(noon)$, $a\sigma_S(noon)$ and $a\sigma(noon)$; $a\sigma_N(dusk)$, $a\sigma_S(dusk)$ and $a\sigma(dusk)$; and $a\sigma_N(midn)$, $a\sigma_S(midn)$ and $a\sigma(midn)$, for the 6-hour MLT sectors around 06hrs, 12hrs, 18hrs and 00hrs, respectively. The first paper in this series (*Lockwood et al.*, 2020a) compared the semi-annual variations in the four $a\sigma$ indices to those in other geomagnetic indices and showed that they revealed a great many of the same characteristics as $am$. The amplification of the semi-annual variation, with respect to that in the estimated power input into the magnetosphere, $P_\alpha$, was shown to increase with distance away from noon, being minimal for the index for the 6-hour sector around magnetic noon, $a\sigma(noon)$, by a factor of near 2 for the $a\sigma(dawn)$ and $a\sigma(dusk)$ (and for the equivalent overall global index $am$) and by a factor of near 3 for $a\sigma(midnight)$.

Throughout this paper we estimate power input into the magnetosphere, $P_\alpha$, using the theoretical estimate devised by *Vasyluinas et al.* (1982). This coupling function is explained, discussed and its use justified at the start of section 3-ii.

### 1-i. *Universal Time variations in different geomagnetic indices*

As shown in Paper 1, the *UT* variation in geomagnetic data is a highly persistent phenomenon. Figure 1 shows average values of various geomagnetic indices in a *UT*-year spectrogram format. The longest data sequence is the homogenised $aa$ index, $aa_H$, generated by *Lockwood et al.* (2018a; b). This index is based on data from just two stations, roughly 180 degrees apart in longitude and so is far from ideal for detecting a *UT* variation. However, $aa_H$ has been compiled using the same model of the stations' sensitivity that was used by *Lockwood et al.* (2019a) and this allows $aa_H$ to capture both the equinoctial pattern and the *UT* variation seen simultaneously in the $am$ index after 1959. *Lockwood et al.* (2018b) show that the equinoctial pattern is present in the $aa_H$ data before the start of the $am$ data in 1959, right back to the start of the $aa_H$ data in 1868. Figure 1f shows that the *UT* variation is also present in $aa_H$ all years before 1959 and appears it was even of larger amplitude before 1930 than in recent decades, although the use of just two stations means that we must use these data with caution in this respect. The $am$ index (Figure 1e) shows a very similar *UT* variation. For $am$ we have hemispheric sub-indices $an$ and $as$ (shown in Figures 1c and Figure 1d) and they both show *UT* variations, but these are almost in antiphase with the peak around 12 UT in $an$ when $as$ is a minimum. Figure 1b shows the well-known strong *UT* variation in the *AL* index. This peaks around the same time as the northern hemisphere mid-latitude indices. Figure 1a shows that the same variation is seen in the northern-hemisphere *SML* index which demonstrates that the *UT* variation in *AL* is not caused by the longitudinal distribution of station locations. There have been attempts to construct an equivalent network to construct southern hemisphere *AE* indices, with limited success because much of the southern hemisphere auroral oval is over sea or ocean and because only relatively short data sequences are available (*Maclennan et al.*, 1991; *Weygand*, 2014). The results of *Maclennan et al.* (1991) clearly showed the antiphase *UT* variation in the southern hemisphere index that is seen in Figure 1 for *as*. The results of *Weygand* (2014) show the same feature, but the amplitudes of the north-south differences are considerably smaller than found by *Maclennan et al.* (1991). Note that both of these studies lacked stations at the key longitudes in the southern hemisphere: between Mawson (MAW) at 62.9°E and W Antarctic Ice Sheet Divide (WSD) at 247.1°E, the only available station is Macquarie Island (MCQ), south of New Zealand at 159.0°E. Also, both studies used the *AE* rather than *AL* (and so reflect some influence of the dayside directly-driven currents detected by *AU*). Not shown in Figure 1 are the polar cap indices *PCN* and *PCS* compiled from magnetometer data from single stations at Thule and Vostock, respectively (*Troshichev et al.*, 2006). The northern hemisphere polar cap index *PCN* (available from 1975 onwards) persistently shows the same *UT* variation as the other northern hemisphere indices shown in Figure 1; however, the southern hemisphere index *PCS* (available for most years after 1995 but only in provisional form) does not show any persistent *UT* variation. The $aa_{HN}$, $aa_{HS}$, *PCN*, and *PCS* indices all come from just one station and so even for the polar cap indices from stations near the centre of the polar cap, the *UT* variation is convolved with local time variations and so variations in photoionization-induced ionospheric conductivity variations. Later in the current paper, we will present, for the first time, the *UT* variations in the hemispheric sub-indices of the four $a\sigma$ indices.

Because they have the most regular network of observing stations, the most reliable data on the *UT* variations in the northern and southern hemisphere are undoubtedly from *an* and *as*, the northern and southern hemisphere sub-indices of *am* (shown in Figure 1c and 1d). The top panel of Figure 2 shows the average *UT* variations of *an*, *as* and *am* for 1995-2017 (dot-dash lines) and for all the available data (for 1959-2019, solid lines). The shorter interval is chosen here because it gives an availability of simultaneous interplanetary data which results in near-continuous estimates of the power into the magnetosphere (values that are accurate to within 5% are available for over 90% of the time: see Figure 3 of *Lockwood et al.*, 2019b). It can be seen that the values of all three indices for the whole interval are considerably higher than for the post-1995 data, which is due to the long-term decline in solar activity that began around 1985. However, the form of the variations is very similar for the two intervals: this is stressed in the lower panel of Figure 2 which shows the variations of the values normalised to the overall mean for the interval (i.e. $an/<an>_{all}$ in red, $as/<as>_{all}$ in blue and $am/<am>_{all}$ in black. It can be seen that the variations for the two intervals in these normalised values are not identical, but they are similar. Figure 2 shows that the hemispheric differences are more complex than a simple antiphase variation with a persistent minimum in both hemispheric indices (and therefore also in $am$) at around 05 $UT$.

Paper 1 and Paper 2 of this series (*Lockwood et al.*, 2020a; b) showed that sorting the $am$ and $a\sigma$ indices by the prevailing polarity of the IMF Y-component in the GSEQ (Geocentric Solar Equatorial) reference frame ($[B_Y]_{GSEQ}$) revealed that the R-M effect is at work, even though these indices show the equinoctial pattern with time of year $F$ and $UT$, rather the pattern predicted for the R-M effect. Figures 3b and 3c show the $UT$ variations for the 1995-2019 interval, but with the data sorted according to the polarity of $[B_Y]_{GSEQ}$, averaged over the prior hour to be consistent with the optimum lag found by *Lockwood et al.* (2019b). The green and mauve lines in Figure 3a show the $UT$ variations of normalised power input to the magnetosphere, $P_\alpha/P_o$ and reveal the average $UT$ variations predicted by the R-M effect (note that the $[B_Y]_{GSEQ} > 0$ data are dominated by enhancements at the March equinox and the $[B_Y]_{GSEQ} < 0$ data are dominated by enhancements at the September equinox as also predicted by the R-M effect). The normalisation is achieved by dividing by $P_o$, the average of $P_\alpha$ for the whole interval (1995-2018) which cancels various constants in the expression for $P_\alpha$. It can be see that all three indices ($an$, $as$, and $am$) reflect the $UT$ variation predicted by the R-M effect in $P_\alpha/P_o$ but there are additional effects, with the $an$ index enhanced around 12 $UT$ for both IMF $[B_Y]_{GSEQ}$ polarities and the $as$ index enhanced around 00 $UT$ for both IMF $[B_Y]_{GSEQ}$ polarities and the $an$ and $as$ indices are both somewhat lower than expected at about 3-8 $UT$. For both IMF $[B_Y]_{GSEQ}$ polarities, $an$ and $as$ (and so, by definition, $am$) are the same around 06 and 18 UT for both IMF $[B_Y]_{GSEQ}$ polarities. The differences between and $an$ and $as$ in Figure 3 are often attributed to hemispheric conductivity differences, and these are indeed a factor, but there is a much larger and more significant factor that is discussed in this paper for the first time in the next section.

**1-ii.** *Motions of the poles and polar caps*

Ionospheric polar cap phenomena are usually ordered and plotted in a geomagnetic coordinate system, for example a geomagnetic latitude and magnetic local time (MLT) system. The pole of these coordinate systems is based on a model of the geomagnetic field and different models assign different geomagnetic coordinates to a given geographic coordinates. There are also different definitions of the magnetic poles to consider: for example, one can use the geomagnetic poles from a fitted dipole (which could be a geocentric dipole for which the dipole axis passes through the centre of the Earth or an eccentric dipole for which, in general, it does not), or the dip pole (where the surface field is vertical). The dip poles in the northern and southern hemisphere have behaved very differently over the last century (*Thébault et al.,* 2015). As shown by the orange points in Figure 4, the northern dip pole has migrated toward the rotational pole such that their separation in geographic latitude of 20° in 1900 has reduced to just 4° in 2020, whereas the southern dip pole has migrated away from the rotational pole such that their separation increased from 18° to 26° in the same interval. Furthermore, the very high latitude of the northern dip pole has allowed the geographic longitudinal separation of the two dip poles to drop from 165° to just 65°. On the other hand, a geocentric dipole model forces the two poles to be 180° apart in longitude and, as shown by the blue points in Figure 4, the (poleward) migration of the two magnetic poles in the same interval is the same and relatively minor (by of order 2.5°). In this paper, we are interested in asymmetries between the two hemispheres and so use an eccentric dipole field model, which again employs a dipole field but does not constrain the dipole's axis to pass through the centre of the Earth (*Fraser-Smith*, 1987; *Koochak and Fraser-Smith*, 2017). This introduces a third type of magnetic pole, namely the eccentric axial poles which is where the fitted eccentric dipole axis threads the Earth's surface. These poles are not, in general, at same latitude nor are they axiomatically 180° apart in longitude. We compute the position of these poles using the equations and coefficients of *Koochak and Fraser-Smith* (2017) which are available for after 1980-2015 and are plotted as the mauve points for 1980, 2000 and 2020 in Figure 4. The northern eccentric dipole axial pole migrated from 8.2° to 5.4° (by 2.8°) from the rotational pole over 1980-2040: in the same interval the corresponding values for the dip pole were 13.1° to 3.5° (by 9.6°) and for the geocentric dipole geomagnetic pole were 11.1° to 9.3° (by 1.8°). In the same interval, the southern eccentric dipole axial pole migrated from 15.3° to 14.4° (by 0.9°) from the rotational pole and the corresponding values for the dip pole were 24.6° to 25.9° (by -1.3°) and for the geocentric dipole geomagnetic pole were 11.1° to 9.3° (by 1.8°, same as for the northern hemisphere). Hence the eccentric poles reflect some of the behaviour of the dip poles, but the changes were considerable smaller, as for the geocentric dipole. The key point that we focus on here is that the offset from the geographic pole for the southern hemisphere exceeds that for the northern hemisphere by a factor of about 2 or more, except for the geocentric dipole for which it is necessarily unity. This ratio of the offsets increases from 1.9 to 2.6 for the eccentric dipole and from 1.9 to 7.4 for the dip poles (largely because the dip pole in the Northern hemisphere has moved so close to the rotational pole). *Koochak and Fraser-Smith* point out that eccentric dipoles have not been exploited in magnetospheric physics despite the obvious importance to hemispheric effects and hence *UT* effects.

A key point about the offset of the magnetic and geographic poles is that it causes motion of the ionospheric footpoints of magnetic field lines toward and away from the Sun. The geographic poles only move very slowly towards or away from the Sun: the orbital motion and elliptical nature of Earth's orbit means that over each the year the poles move together toward and the away from the Sun but with a peak velocity of only 0.5 ms$^{-1}$. However, the diurnal rotation of the magnetic poles around the geographic poles yields a considerably faster velocity toward and away from the Sun, which increases linearly with the offset in geographic latitude of the magnetic pole from the rotational pole. The expected Universal Time effect of this on the auroral oval, and on the open field line polar cap inside it, has been described using an empirical model by *Tsyganenko* (2019) who showed that there is only minor distortion of the shape of the oval such that the circular motion of the oval in GSEQ (Geocentric Solar Equatorial) *XY* plane largely reflects that of the geomagnetic pole. Observationally, *Newell and Meng* (1989) surveyed 3 years' data from the DMSP (Defense Meteorological Satellite Program) F7 satellite and showed that the region of cusp precipitation migrated in geomagnetic latitude by about 0.06° for each 1° shift in dipole tilt angle. That means that 94% of the motion of the magnetic pole in the GSEQ frame is reflected in the cusp location and only 0.6% in the geomagnetic frame. The cusp precipitation is on newly-opened field lines generated by magnetopause reconnection (*Smith and Lockwood*, 1996) and hence the motion of the dayside open-closed boundary in GSEQ largely reflects that in the magnetic pole. *Vorobjev and Yagodkina* (2010) showed that the magnetic latitude of the poleward edge of the nightside northern-hemisphere auroral oval, as detected in DMSP satellite data from 1986, was shown to have a sinusoidal diurnal variation in amplitude near 2°, whereas the offset of the rotational northern eccentric axial pole at that time was about 8°. Hence in the GSEQ frame only about 75% the motion in the magnetic pole is reflected in this boundary. However, this boundary is generally equatorward of the open-closed field line boundary (OCB) and this is likely to make this percentage a poor estimate of the real value for the nightside OCB. The OCB can be identified in global MHD simulations and *Kabin et al.* (2004) found that magnetic latitude shifts in the noon OCB were 1.3° and −0.9° for dipole tilts of +35° and −35°, i.e., 3.9% and 2.7%, respectively, consistent with the results on the cusp by *Newell and Meng* (1989). The corresponding shifts in the midnight OCB were 0.8° and −0.5° (2.3% and 1.5%, respectively). Hence these simulations show the nightside OCB moves even more closely with the magnetic pole than the dayside OCB. The fact that the OCB is largely moving with the geomagnetic poles shows that closed field lines outside the open field line region are also taking part in this diurnal wobble caused by the pole motion.

*Oznovich et al.* (1993) showed that during low auroral activity the auroral oval as a whole was shifted by 1 degree in geomagnetic coordinates for every 10-degree change in the dipole tilt angle. This yields an estimate that 90% of the motion of the geomagnetic pole in the GSEQ frame induced by the diurnal motion of the pole is reflected in the auroral oval as a whole. Being at large longitudinal separations (if not exactly the 180° for a geocentric dipole model) the motion of the auroral ovals induced by the magnetic pole motions would be in close to, but not exactly, in antiphase in the GSEQ frame with the southern pole moving antisunward when the northern is moving sunward, and vice-versa. This has been directly observed by *Stubbs et al.* (2005) using full and simultaneous auroral images of the northern and southern auroral ovals made by the IMAGE and Polar satellites. These images are here reproduced in Figure 5, where the auroral intensity is plotted in a geomagnetic latitude-magnetic local time (MLT) frame: the altitude-adjusted corrected geomagnetic (AACGM) coordinate system was used (*Baker and Wing*, 1989). The white dots show the geographic poles which are points that are essentially fixed the GSEQ frame, their motion due to Earth's annual orbit being very slow. *Stubbs et al.* fitted circular polar cap boundaries to the poleward edge of the northern and southern auroral oval in the geomagnetic latitude-MLT frame by varying the radius and centre location in the noon-midnight and dawn-dusk directions and Figure 6 shows those boundaries mapped into the Geocentric Solar Equatorial frame (GSEQ) where $X_{GSEQ}$ points from the centre of the Earth to the centre of the Sun and $Y_{GSEQ}$ lies parallel to the solar equatorial plane and points broadly from dusk to dawn. (Details of the mapping procedure are given in Section 2). Also shown as a cross is the corresponding location of the eccentric dipole axial pole. The oval and pole are shown for three times half an hour apart (11:20 in green, 11:50 in orange and 12:20 in mauve). We use this interval because *Stubbs et al*. show that during it the radius of the two polar caps was increasing but only very slightly, which makes the migration of the polar cap easier to discern because it is not complicated by expansion or contraction as it moves. Part (a) is for the northern hemisphere oval and shows both the pole and the oval moving toward the Sun; (b) is for the southern hemisphere oval and shows both the pole and the oval moving away from the Sun and in the $-Y_{GSEQ}$ direction. The oval moves as a whole with motion that closely corresponds to that of the eccentric dipole pole, as expected from the above discussion.

## 1-ii. *Effect of solar wind dynamic pressure*

Paper 2 in this series (*Lockwood et al*., 2020b) reviews past studies revealing an independent effect of solar wind dynamic pressure on geomagnetic activity. Paper 2 shows that the geomagnetic response to a given injected power into the magnetosphere is increased if the solar wind dynamic pressure $p_{SW}$ is increased. Furthermore, Paper 2 also shows that the amplitude of the equinoctial pattern increases with increased dynamic pressure as does the amplitude of the *UT* variation. Using models, Paper 3 (*Lockwood et al.,* 2020c) has shown that a good explanation of this was the effect of dynamic pressure squeezing the tail and increasing both the energy stored in the near-Earth tail and the current in the cross-tail current sheet. This idea had been proposed by *Lockwood* (2013) as an explanation of why the equinoctial pattern was seen in association with the substorm current wedge and why it has a dependence on the square of the solar wind velocity (*Finch et al*., 2008). The modelling in Paper 3 indicates that the dipole tilt changes the ability of the solar wind dynamic pressure to modulate both the energy stored and the cross-tail current and that hemispheric asymmetry in the field means that positive dipole tilts have different effects to negative dipole tilts, thereby introducing a *UT* variation.

Figure 1 of Paper 3 (*Lockwood et al*., 2020c) shows that geomagnetic activity (quantified by the *am* index) is enhanced at constant power input to the magnetosphere (which depends on IMF orientation) by enhanced solar wind pressure, $p_{SW}$ (which does not depend on IMF orientation). Another plot showing the relationship of the effects of dynamic pressure and power input on *am* is Figure 19 of Paper 2 (*Lockwood et al.*, 2020b). The independent effect of $p_{SW}$ also supported by the modelling shown in Paper 3 which shows that, for a given magnetospheric open flux (that depends on the prior history of the IMF orientation), energy stored in the near-Earth tail lobes and cross-tail current are both increased by enhanced solar wind dynamic pressure, and both of have the potential to enhance geomagnetic activity. This evidence is brought together, using different plots, and summarised in Appendix A to the present paper. Figure 7 presents further details of the effect of solar wind dynamic pressure using all *am* index and solar wind data for 1995-2019 (inclusive). The *am* data are interpolated linearly to 1-minute values from the 3-hourly raw index data and compared to 1-minute interplanetary data allowing for the 60 minute response lag found in Paper 2. There are 3 groups of four panels in Figure 7: the top 4 (a-d) are for the simultaneous $p_{SW}$ (allowing for the 60 minute lag) in the lower tercile of its overall occurrence distribution, $p_{SW} < q(0.33)$, where $q(0.33)$ is the 1/3 quantile of the cumulative distribution of all 1-minute $p_{SW}$ values, the middle 4 panels (e-h) are for the middle tercile of this near-simultaneous solar wind dynamic pressure, $q(0.33) \leq p_{SW} < q(0.67)$ and the lower 4 panels (i-l) are for the upper tercile of the near-simultaneous solar wind dynamic pressure, $p_{SW} \geq q(0.67)$. The data are also sorted according to the polarity of the Y-component (in the GSEQ frame) of the IMF, with the left-hand panels being for IMF $[B_Y]_{GSEQ} < 0$, and the right-hand panels being for IMF $[B_Y]_{GSEQ} > 0$. The panels are organized in pairs with the upper plot of each pair showing the probability distribution function (p.d.f.s) of normalised power input to the magnetosphere, $P_\alpha/P_o$, as a function of fraction of the year $F$ and the lower of each pair showing the corresponding plot of the normalized *am* amplification factor, $[am/P_\alpha]_n = (am/<am>_{all})/(P_\alpha/P_o)$, as a function of $F$ and in the same $F$ and $P_\alpha/P_o$ bins as the p.d.f.s in the plot above it. The upper plots all show the Russell-McPherron (R-M) is at work, with normalised power input to the magnetosphere, $P_\alpha/P_o$, increased around the "favoured" equinox, which is the March equinox (around $F = 0.22$) for $[B_Y]_{GSEQ} < 0$ and the September equinox (around $F = 0.73$) for $[B_Y]_{GSEQ} > 0$. Because there are some common factors in the expressions for $P_\alpha$ and $p_{SW}$ (specifically, the solar wind speed $V_{SW}$, mean ion mass $m_{SW}$, and number density $N_{SW}$) larger values of $P_\alpha/P_o$ are more common for larger $p_{SW}$ (see Figure 19 of Paper 2). The lower panels in each pair show the level of the *am* response for unit power input to the magnetosphere in the same bins of $F$ and $P_\alpha/P_o$ as the upper panels and so show the amplification factor of *am*. This is greatest around the equinoxes and increases with $p_{SW}$. Hence there is a clear amplification of $am$ at the equinoxes that depends on $p_{SW}$ but is an independent effect from the R-M effect. Appendix A shows two new plots that summarise findings presented in paper 1 and two that stress how important the effect of $p_{SW}$ is to the generation of the semi-annual variation.

### 1-iii. *The equinoctial and Russell-McPherron time-of-year/time-of-day patterns*

Figure 8c compares the theoretical Russell-McPherron and equinoctial effects by overlaying the $F$-$UT$ patterns, both derived using the eccentric dipole geomagnetic field model of *Koochak and Fraser-Smith* (2017) with constants interpolated to the year 2007 (the mid-point of 1995-2019, the interval for which 1-minute resolution interplanetary data are available). The dipole tilt angle $\psi$ was computed as a function of $F$ and $UT$, being the angle between the Earth's (eccentric) dipole axis $\vec{M}$ and the geocentric position vector of the subsolar point, $\vec{S}$, computed using the SUBSOL routine of the LOWTRAN7 Sun and Moon Models Matlab software package. The colour contours in Figure 8c give the absolute value of the dipole tilt angle, $|\psi|$, superposed on which are contours showing the IMF orientation factor used by $P_\alpha$, namely $A_\theta = sin^4(\theta/2)$, where $\theta$ is the clock angle of the IMF in the GSM frame, computed from a given IMF orientation in the GSEQ frame using the CXFORM Coordinate transformation package described above. These predictions are the average for an equal mix of $[B_Y]_{GSEQ} = -|B| < 0$ and $[B_Y]_{GSEQ} = +|B| > 0$ and contours are shown for $< A_\theta >$ of 0.28 (in black) and 0.31 (in mauve). The $< A_\theta > = 0.31$ contours in Figure 8c define the two Russell-McPherron peaks and are used in the derivation of parts a and b. These show the mean values of $|\psi|$ over the range of $F$ defined by the maximum extent in $F$ of the mauve $< A_\theta > = 0.31$ contour as a function of $UT$: Figure 8a is for the March equinox and Figure 8b for the September equinox. The area shaded pink is the $UT$ extent of the peak defined by the corresponding mauve contour in Figure 8c. Time increases up the plots and parts a and b which show that during and after the peak in $< A_\theta >$ (and hence also in $< P_\alpha >$) the variation in $< |\psi| >$ is identical for the two equinoxes and hence no asymmetry is introduced between them and there is no net $UT$ variation when they are averaged together. This conclusion was found to hold for both geocentric and eccentric dipole fields and all epochs. This means that asymmetry between the March and September peaks, and hence a $UT$ variation, is not introduced into either predicted patterns (nor into the relationship between the two), by the magnetic field model as long as the field has a single dipole axis, even if it is an eccentric one that does not pass through the Earth's centre.

**1-iv.** *Aims of this paper*

In Section 2 of this paper, we investigate the effect of the motion of the open polar caps on the $UT$ dependence of geomagnetic activity using the eccentric dipole model of the geomagnetic field of *Koochak and Fraser-Smith* (2017). In section 3 we discuss how we model averages of mid-latitude "range" geomagnetic indices studied in this paper ($am$ and its hemispheric sub-indices, $an$ and $as$, the four $a\sigma$ indices and the two hemispheric sub-indices of each) as a function of $F$ and $UT$. This involves developing factors that allow for the polar cap motions discussed in section 2, for ionospheric conductivities, for the R-M effect in solar wind forcing and for the effect of dynamic pressure and dipole tilt on the near-Earth tail. In Section 4 we compare the modelled $F$-$UT$ patterns with those for conductivity-corrected versions of the $am$, $an$ and $as$ indices, and in section 5 we do the same for all 12 of the hemispheric and global $a\sigma$ indices to study how the model performs in the four 6-hour MLT sectors. In section 6 we also apply the model to one example (the midnight $a\sigma$ index) broken down into two subsets of the prevailing IMF Y-component. Sections 3-6 all deal with modelling the average values of the indices (at a given $F$ and $UT$) and in section 7 we look at large and near-extreme events in the $am$ index and show the modelling has implications, which, in itself, raises interesting questions as to why and how. Section 8 contains a summary discussion and conclusions.

## 2. Effect of diurnal pole motions

In this section we consider the effect of the daily motions of the magnetic poles due to Earth's rotation. For an axisymmetric, geocentric dipole field, the motions of the two poles would be equal and opposite and any nett global effect would not show any $UT$ variation. To allow for the large (and currently increasing) asymmetry in the geomagnetic field, we here use the eccentric dipole field of *Koochak and Fraser-Smith* (2017). Results are presented for the interpolated eccentric dipole coefficients that apply to the years 1989 (the midpoint of the 1959-2019 interval of all $am$ data), and 2007 (the midpoint of the 1995-2019 interval of quasi-continuous interplanetary data) but were also generated for 1980 and 2015 and are only different in small details for our purposes.

As for Figure 6, we mapped the location of the eccentric axial poles in geographic coordinates into the GSEQ frame using the CXFORM Coordinate transformation package, initially written by Ed Santiago of Los Alamos National Laboratory and Ryan Boller of NASA's Goddard Space Flight Centre and re-coded for Matlab by Patrik Forssén (SatStar Ltd & Karlstad University) in 2017. This software package is based on the equations by Mike Hapgood of RAL Space, Rutherford Appleton Laboratory (*Hapgood*, 1992).

The loci of the eccentric dipole axial poles in the $XY$ frame of the GSEQ reference frame are shown in Figure 9, in which the rotation of the Earth makes the poles rotate clockwise. The points show the location of the poles at 12 $UT$. The greater offset in the axial eccentric dipole poles from the rotational pole in the southern hemisphere makes the radius of the orbits larger for the southern hemisphere. Figure 10 shows the $X$-component (sunward) velocity of the two poles in the GSEQ frame as a function of $UT$. The two are in close to antiphase (but not exactly as the longitudinal separation of the axial poles is not 180°) and the larger offset of the pole in the southern hemisphere means that the amplitude of its sinusoidal variation in the velocity in the southern hemisphere, $V_{XS}$, is larger than that for the northern hemisphere, $V_{XN}$. The open field line region in one hemisphere moves as a whole in GSEQ because the geomagnetic field, both open and closed field lines, moves as a whole. There appears to be some $UT$-dependent distortion of the region of open flux, presumably induced by changing pressure balance between open and closed field lines, because observations of dipole tilt effects on locations of the open-closed boundary in geomagnetic reference frames, as discussed in section 1–i, reveal that up to 10% of the dipole tilt variation is reflected in the motion in the inferred or modelled OCB boundaries in a geomagnetic frame. This means that at least 90% of the diurnal motion of the magnetic pole and the geomagnetic frame in the GSEQ frame, and in particular its motion toward and away from the Sun, must be reflected in motion of the open field line region, as a whole, in GSEQ.

Figure 11 illustrates why this has an influence. The Expanding-Contracting Polar Cap (ECPC) model of the excitation of ionospheric convection in non-steady-state situations (*Cowley and Lockwood*, 1992) is based on the fact that only in steady-state (or for averages over sufficient timescales that steady state applies) does the solar wind electric field map down open field lines into the polar ionosphere: by Faraday's law (in differential form) a non-zero rate of change of magnetic field $B_{TL}$ in the tail lobe, on open field lines between the solar wind and the ionosphere gives a curl of the electric field (whereas in steady state $\nabla \times \vec{E} = -\partial \vec{B}/\partial t = 0$) and, integrated down the open field lines, this decouples the electric field in the polar cap from that in interplanetary space (i.e., there are induction effects). Figure 11 is a schematic based on that by *Lockwood and Cowley* (1992) and *Lockwood and Morley* (2004) that shows open field lines mapping from the ends of reconnection X-lines AB in the dayside magnetopause and DE in the cross-tail current sheet, mapping down open field lines on the open-closed boundary to their ionospheric footpoints, the ends of the "merging gaps", ab and de, respectively. To look at the total decoupling of, for example, the voltage $\Phi_{CF}$ across the "Stern Gap" CF (the region of open magnetospheric field lines in interplanetary space such that $\Phi_{CF}$ is the integral of the interplanetary electric field along the line CF) and that along the ionospheric polar cap diameter (the "transpolar voltage" or "cross-cap potential drop", $\Phi_{cf}$) we use Faraday's law in integral form by considering the loop CFfc and neglecting any field-parallel voltages along the field lines Cc and Ff (we know these to be comparatively very small from the minimum energies of primary precipitating electrons or ions):

$$\oint_{CFfc} \vec{E}.\overrightarrow{dl} \; = \; \Phi_{CF} + \Phi_{fc} \; = \; \Phi_{CF} - \Phi_{cf} \; = \; -\frac{\partial}{\partial t} \int_{CFfc} \vec{B}.\overrightarrow{dA} \qquad (1)$$

hence the decoupling is caused by a change in the total magnetic flux threading the loop. The same applies to the nightside reconnection voltage $\Phi_{DE}$ (the integral of the reconnection rate) and the loop DEed and the magnetopause reconnection voltage $\Phi_{AB}$ and the loop ABba. When using equation (1) it is important that $\vec{E}.\overrightarrow{dl}$ is evaluated by moving around the loop in a common, right-hand sense and that for all parts of the loop the electric field is quantified in a common frame of reference (i.e., it is a "fixed loop"). We here use the geocentric GSEQ frame which shares the same sunward X axis as all the frames in which we measure the solar wind speed and hence interplanetary electric fields and voltages (such as GSE and GSM), and in which have shown (Figures 9 and 10) that the footpoints ab, cf and de have a sinusoidal velocity variation in the *X* direction. If we use the estimate that 90% of the variation for the axial pole is reflected in the open-closed boundary (see section 1-ii), from Figure 10 we find this polar motion velocity in the X-direction $[V_X]_{PM}$ has a sinusoidal variation of amplitude of about 100 ms[-1] for the southern polar cap and about 50 ms[-1] for the northern at all times of year. Using an ionospheric magnetic field strength of $\vec{B}_i$ of 4.5×10[-5]T in the topside ionosphere in the $-Z$ direction, this gives an electric field $\vec{E}_{PM} \; = \; -\vec{V}_{PM} \times \vec{B}_i$ with a dawn-dusk component $E_{PM} \approx B_i [V_X]_{PM}$ that has sinusoidal *UT* variations of amplitudes 5 mVm[-1] and 2.5

mVm$^{-1}$ for the southern and northern hemispheres, respectively. This also applies to the merging gaps ab and de as well as the polar cap diameter cf because, to a first approximation, the polar cap moves as a whole. A useful comparison is with the electric field in interplanetary space: in the GSEQ frame the solar wind speed is typically 400 kms$^{-1}$ which for a flow-transverse IMF component of 5nT is an interplanetary electric field of $E_{SW} \approx 2$ mVm$^{-1}$. However, the best way to put these $UT$ variations into context is to consider the magnitude of their effect, relative to the transpolar voltage $\Phi_{cf}$: a circular polar cap of angular radius 15° gives a polar cap area of $A_{pc} \approx 10^{13}$ m$^2$, an open flux of $B_i A_{pc} \approx 5 \times 10^8$ Wb and a polar cap diameter at an altitude of 850 km of $d_{pc} \approx 3780$ km, for which the southern hemisphere pole motion gives a sinusoidal $UT$ voltage modulation of amplitude $\Delta\Phi_{PM} = d_{pc} E_{PM} \approx \pm 19.0$ kV and for the northern hemisphere pole motion gives ±9.5 kV.  These are not negligible fractions of typical values of $\Phi_{cf}$: for example, *Lockwood et al.* (2009) find that average values of $\Phi_{cf}$ during quiet times (when the polar cap flux of $5 \times 10^8$ Wb is appropriate) is 25kV. This rises to 52 kV during substorm growth phases, 64 kV between substorm onset and peak expansion, 72 kV between peak expansion and the start of recovery, 67kV in substorm recovery phases and 83 kV during steady convection events.  *Lockwood et al.* (1990) find the polar cap flux increases to about $10^9$ Wb during steady convection events which increases the polar cap radius and hence the predicted $UT$ variations due to pole motion by a factor $2^{1/2} \approx 1.4$.  Hence for the northern hemisphere the percentage southern hemisphere $UT$ induction effect is, on average, $100 \times \Delta\Phi_{PM}/\Phi_{cf} \approx \pm 76\%$ in quiet times, falling to about ±32% during steady convection events.   Values for the northern polar cap are roughly half of these.  When a polar cap is moving sunward it is effectively adding to the effect on open field lines of the (antisunward) solar wind flow and in the other half of the diurnal cycle, when is moving away from the Sun, it is reducing the effect of the solar wind flow.

From equation (1) this will have a mixture of two effects. The direct effect would be the modulation of the transpolar voltage $\Phi_{cf}$ by the diurnal motion of the magnetic pole for given solar wind electric field and Stern Gap voltage $\Phi_{CF}$.  If this were the only effect, then the change in observed transpolar voltage would equal the voltage induced by the polar cap motion so $\Delta\Phi_{cf} = \Delta\Phi_{PM}$. However in general we should not expect the full effect of the pole motion to appear in the transpolar voltage and in general

$$\Delta \Phi_{cf} = c_{PM} \Delta \Phi_{PM} , \qquad \text{where } 0 \leq c_{PM} \leq 1 \qquad\qquad (2)$$

There are two reasons for this. The first is the "flywheel effect" of thermospheric inertia whereby collisions between ionospheric ions, particularly in the E-region, and the (much) more numerous neutral thermospheric atoms and molecules tend to keep ions moving at the same speed even if the solar wind forcing of ionospheric convection changes (*Deng et al.*, 1993). However, a more intrinsic cause of the factor $c_{PM}$ is predicted by the ECPC model of ionospheric convection excitation (*Cowley and Lockwood*, 1992). The ionospheric polar motions have no effect on the conditions at the magnetopause and cross-tail reconnection sites and so do not directly modulate the voltages $\Phi_{AB}$ and $\Phi_{DE}$ with which open field lines are opened and closed, respectively. In the ECPC model of ionospheric convection excitation, opening and closing of field lines perturbs the location of the open-closed field line boundary in the ionosphere and ionospheric convection is the response of the ionospheric flows as the boundary tends towards the new equilibrium configuration. This might appear to argue that the pole motions gave no effect on ionospheric flows (and hence $c_{PM} = 0$) but this overlooks the fact that the motion of the polar cap will also influence the equilibrium configuration that the system relaxing back toward.

Hence, in the limit $c_{PM} = 1$ the pole motions induce only a directly driven response in polar cap flows and in the limit $c_{PM} = 0$ they have no effect of the ionospheric flows and only generate inductive changes in the tail lobe field, i.e., the response is of a purely storage/release nature. The general value of $c_{PM}$ between these two limits is a mixture of both effects. For general $c_{PM}$, the change in observed transpolar voltage $\Delta \Phi_{cf}$ is less than $\Delta \Phi_{PM}$ but is not zero. This means that, as well as modulating ionospheric voltages directly, the pole motions would modulate the rate of energy storage or release in the tail lobe: additional energy would be stored when the pole is moving sunward, and this would be released again 12 hours later when it is moving antisunward. Because the poles are close to being 180° of geographic longitude apart, motions in the two hemispheres are close to being in antiphase and hence the energy in one tail lobe grows while the other declines but because the of hemispheric asymmetry in the geomagnetic field, these *UT* variations do not cancel.

Figure 10a predicts the pole motion effect will be roughly double the size in the southern hemisphere to the northern and will peak at 12 *UT* in the northern hemisphere (with a minimum at 0 $UT$) and at 22 $UT$ in the southern hemisphere (with a minimum at 10 $UT$). From Figure 10b a global effect that is the average of that for both hemispheres will peak near 21 $UT$ with a minimum near 9 $UT$. There are strong elements of these predicted $UT$ variations seen in the $an$, $as$, and $am$ indices shown in Figure 2, but they are clearly also modulated by other factors. This is not surprising as we know that there are other consistent variations with $UT$ (at a given time of year), such as the effect of ionospheric conductivities, the Russell-McPherron effect and the squeezing of the tail by solar wind dynamic pressure. These all occur concurrently and in the remainder of this paper we investigate how the pole motion effects described in this section interact with other factors.

Note that in the modelling of geomagnetic indices presented in the following section, all parameters can be quantified using approximations and/or averages of observations – with one exception: there is no way to quantify the parameter $c_{MP}$. The closest we could get would be a global MHD numerical model, but this would be far from ideal because the lower boundary is not in the ionosphere and tends to be at an altitude of about $3R_E$ for computational reasons. Hence including ionospheric pole motions in a self-consistent way will not be a straightforward task. As a result, we have to treat $c_{MP}$ as a free fit parameter. There are other parameters that are quantified by fitting to data. A factor $c_\psi$ is needed to relate the effects of the magnetic shear across the near-Earth tail current sheet (i.e., the current in that sheet) to its effect on the geomagnetic index in question, and this can be quantified directly by studying the index response to modelled changes in that current. Similarly, we use a factor $c_{RM}$ to scale the theoretical R-M forcing pattern to the observed pattern of power input into the magnetosphere, $P_\alpha(F, UT)/P_o$: again, this is done by a direct fit to the data.

### 3. Modelling geomagnetic indices.

In the modelling presented here we use 3 multiplicative normalised factors to generate a simulation of a given geomagnetic index (here given the generic name $ax$), $ax_{Hm1}$ (where H signifies the hemisphere it applies to, i.e., N for north and S for south:

$$ax_{Hm1}(F, UT) = <ax_H> . P_{RM}(F, UT). P_\psi(F, UT). [P_{PM}(F, UT)]_H \qquad (3)$$

where the terms $P_{RM}(F, UT)$, $P_\psi(F, UT)$ and $[P_{PM}(F, UT)]_H$ account for the effects on $ax$ of, respectively, the Russell-McPherron effect in solar wind-magnetosphere coupling, dynamic pressure and dipole tilt effects on the tail lobe, and the pole motions, as described in the following subsections. Each of these factors is normalised so its average value over all $F$ and $UT$ is unity and so the modelled pattern is then scaled by multiplying by the average value of the index in question, $<ax_H>$ for the full period of in integer number of years (i.e., over all $F$ and $UT$). This was done for the hemispheric indices, $an$ and $as$, and the $am$ index was then modelled as the average of the two. There is one more factor that is not included in equation (3) namely an allowance for the effect of ionospheric conductivities $P_{\Sigma H}(F, UT)$. This factor is different for each index and in this paper we are dealing with 15 indices. Rather than include $P_{\Sigma H}(F, UT)$ in an equation (3) for each index we here adopt a procedure to remove the conductivity effects first to generate a conductivity-corrected $F$-$UT$ pattern

$$ax_{Hcc}(F, UT) = ax_H(F, UT)/P_{\Sigma H}(F, UT) \qquad (4)$$

and we then model the conductivity-corrected index $ax_{Hcc}$ using equation (3). Because in this procedure the (hemisphere-specific) conductivity correction has already been made (using equation (4) with a factor $P_{\Sigma H}(F, UT)$, the derivation of which is explained in the next section), hemispheric differences between $ax_{Nm}$ and $ax_{Sm}$ will only be due to the pole-motion term that is the main focus of the present paper.

### 3-i. *Allowance for conductivity effects*

With the terms $P_{\Sigma N}(F, UT)$ and $P_{\Sigma S}(F, UT)$, we allow for the effects of both the ionospheric Hall and Pedersen conductivities generated by photoionization. Note that this excludes enhancement of the conductivities over a background level (associated with the quiet auroral oval) by enhanced particle precipitations which will depend on the location (especially in relation to the auroral oval) and the activity level. We regard these enhanced particle precipitation effects on conductivity as an intrinsic part of the activity index that we are modelling. The ionospheric Hall and Pedersen conductivities generated by photoionization both depend on the solar zenith angle (e.g., *Ieda et al.*, 2014) and so, at any fixed geomagnetic location, on the tilt angle $\psi$. In theory, the conductivities could be evaluated for every location in the polar regions using empirical relationships for a given solar zenith angle, $\chi$ and sunspot activity level. However, this leaves the problem because we do not know which locations most influence the index under consideration, it could be the auroral oval, over the stations or a mix of the two (*Lockwood et al.* 2018b, 2019a). Hence, we take an empirical approach using means over several days of the deviations of the northern hemisphere index and simultaneous southern hemisphere index. We then study their variation with time of year $F$ and compare with the means of the dipole tilt angle $\psi$. It is assumed as a first-order approximation that on these timescales the hemispheric differences are due to conductivity effects alone and the deduced variation with $\psi$ caused by changes in $F$ will also apply to the variations with $\psi$ caused by changes in $UT$.

Hence, for example for the $am$, $an$ and $as$ indices we study the dependence of $\Delta an = (an - am)$ and $\Delta as = (as - am)$ with $F$ and compare with the corresponding variations of the mean dipole tilt angle, $\psi$. Figure 12 presents the results. Figure 12a shows the variations with $F$ of the normalised observed indices $am/< am >$ (in black), $an/< an >$ (in red) and $as/< as >$ (in blue). These are averages for all available data that are for 1959-2019, inclusive and hence there are 4951 3-hour samples in each $F$ bin. All three indices show the semi-annual variation clearly, but $an$ and $as$ show the clear effect of photoionization conductivity enhancement with enhanced index values in summer and reduced values in winter in both cases. Figure 12c shows the variations with $F$ of $\Delta an/< an >$ (in red) and $\Delta as/< as >$ (in blue) which are close to being mirror images of each other. The same is true of the variations of these ratios with $\psi$ shown in Figure 12d. The best 4th-order polynomial fit for $\Delta an/< an >$ is:

$$\Delta an/an = 1.064 \times 10^{-8} \psi(F)^4 + 2.840 \times 10^{-6} \psi(F)^3$$
$$+ 1.910 \times 10^{-5} \psi(F)^2 + 1.695 \times 10^{-3} \psi(F) - 0.884 \times 10^{-2} \qquad (6)$$

The choice of polynomial order $n$ was made by measuring the fit residuals as a function of $n$. The r.m.s. fit error decreased with $n$ but going from $n = 4$ to $n = 5$ only decreased it by 0.2%. The concern is that use of too high a value for $n$ would render unrealistic the extrapolations from the largest/smallest $\psi$ datapoints to the largest/smallest possible $\psi$ values. It was found that although $n = 4$ gave the largest gradients $(d\Delta an/d\psi)$ at these extremes it was the largest value of $n$ that gave a 2nd-order derivative $(d^2\Delta an\,/\,d\psi^2)$ that varied close to linearly with $\psi$ at all $\psi$. For a first-order correction we take $am$ to be a good estimate of $an_{cc}$ on the approximately 10-day timescales considered in Figure 12, then

$$an_{cc} = an - \Delta an = an/P_{\Sigma N}(\psi) \qquad (7)$$

hence the northern hemisphere conductivity factor is

$$P_{\Sigma N}(\psi) = (1 - \Delta an/an)^{-1} \qquad (8)$$

The corresponding best 4th-order polynomial fit for the southern hemisphere index $as$ is

$$\Delta as/as = -0.948{\times}10^{-8}\,\psi(F)^4 - 3.137{\times}10^{-6}\,\psi(F)^3$$
$$-2.238{\times}10^{-5}\,\psi(F)^2 - 1.138{\times}10^{-3}\,\psi(F) - 0.988{\times}10^{-2} \qquad (9)$$

and as for the northern hemisphere, the conductivity-corrected $as$ index is

$$as_{cc} = as - \Delta as = as/P_{\Sigma S}(\psi) \qquad (10)$$

and the southern hemisphere conductivity factor is

$$P_{\Sigma S}(\psi) = (1 - \Delta as/as)^{-1} \qquad (11)$$

The corrected indices, $an_{cc} = an/P_{\Sigma N}$, $as_{cc} = as/P_{\Sigma S}$ and $am_{cc} = (an_{cc} + as_{cc})/2$ are shown in Figure 12b. Note that the corrections make $an_{cc}$ and $as_{cc}$ very similar indeed and also that the resulting $am_{cc}$ is not exactly the same as $am$: the semi-annual variation in $am_{cc}$ is slightly larger in amplitude and there is different structure around the peaks (which is also seen in both $an_{cc}$ and $as_{cc}$). This indicates that the conductivity effects in the two hemispheres do not exactly cancel in $am$. The residuals for the polynomial fits give a percentage root mean square (r.m.s.) fit residual error in $P_{\Sigma N}$ and $P_{\Sigma S}$ of just 0.21%.

Figure 13 shows the corresponding plots to Figure 12d (which is for the $am$ index) for the 4 pairs of hemispheric $a\sigma$ indices: (a) $a\sigma(dawn)$; (b) $a\sigma(noon)$; (c) $a\sigma(dusk)$; and $a\sigma(midn)$. It can see that the photoionization conductivity correction is greatest for the noon sector and very small for the midnight sector. Figure 14 shows the resulting time-of-year ($F$) variations of the conductivity-corrected indices and corresponds to Figure 12b. In each case, the corrected index for the northern hemisphere is very similar to that for the southern and the semi-annual variation is clearly seen. Furthermore, the variations for each MLT sector are very similar indeed and similar to that for $an_{cc}$, $as_{cc}$ and $am_{cc}$ shown in Figure 12b: even the small-scale structure around the equinox peaks is the same in each case. The amplitude of the fractional semi-annual variation (as a ratio of the overall mean) is similar in each case, but still smallest for noon and greatest for midnight.

The polynomial fits giving the $P_{\Sigma H}(\psi)$ factors can be used with the computed $\psi(F, UT)$ pattern shown in Figure 8c to compute the $P_{\Sigma H}(F, UT)$ conductivity correction factors for all the hemispheric indices. The results are shown in the Figure 15 and are used to correct the indices for conductivity effects using equations (7) and (10) and the corresponding equations for the $a\sigma$ indices.

### 3-ii. *The Russell-McPherron factor, $P_{RM}$*

The top row in Figure 16 gives the $F$-$UT$ patterns of the power input into the magnetosphere, $P_{\alpha}$ (colour pixels) estimated from 1-minute interplanetary parameters for 1980-2019 (inclusive) and here normalised by dividing by its mean value for all data, $P_o$. This normalisation has the advantage of cancelling various constants in the equation for $P_{\alpha}$ and also removes the need to be repeatedly quoting large absolute power values. This is based on the dimensional analysis theory by *Vasyluinas et al.* (1982) and the derivation is described in *Lockwood* (2019) who shows that $P_{\alpha}/P_o$ correlates very highly with the $am$ index and that the one major limitation in the theoretical formulation of $P_{\alpha}$ (the omission of the relatively small solar wind Poynting flux) causes only very small errors. *Finch and Lockwood* (2007) have shown $P_{\alpha}$ performs better than (or as well as) all other simple coupling functions on all timescales between 3 hours and 1 year (on timescales approaching one year, IMF orientation factors average out and simpler coupling functions perform as well). One feature unique to the *Vasyluinas et al.* (1982) formulation is it employs all the relevant interplanetary variables but has only one free fit parameter, the coupling exponent, α. This is important because ascribing a free fit parameter exponent to each variable considerably increases the danger of "overfitting" whereby a good fit is obtained to the training data that is not sustained in test data because the noise has been fitted. There are considerable sources of noise in solar-wind/magnetosphere coupling studies including measurement errors, propagation lag uncertainties, the fact that solar wind seen by the upstream monitoring spacecraft may not actually hit the Earth and, most of all, data gaps in the data series, which are a particular problem especially if any data from before 1995 are employed in the fitting and training. We here use the criteria to define a valid value of $P_\alpha$ (to a given required accuracy) that were derived by *Lockwood et al.* (2019b). These were derived in a study that introduced synthetic data gaps at random into almost continuous interplanetary data and studied how much they changed the derived values from the known correct value. In the present paper, we only use a $P_\alpha/P_o$ value if the uncertainty due to missing data is estimated to be less than ±5%. All subsequent patterns as a function of fraction of a calendar year ($F$) and Universal Time ($UT$) are generated by averaging the hourly data for each $UT$ in 36 equal width bins of $F$ (each just over 10 days in width). This yields 864 $F$-$UT$ bins and we are applying a 2-dimensonal 1-3-1 triangular weighting smooth in both the $F$ and $UT$ dimensions. For the observations, the geomagnetic index data are linearly interpolated to hourly values from the 3-hourly indices using linear interpolation.

Figure 16a shows the $F$-$UT$ pattern of $P_\alpha/P_o$ for the subset of the data when the mean IMF Y component in the GSEQ frame, $[B_Y]_{GSEQ}$ was negative over the prior hour. Figure 16b shows the corresponding plot for $[B_Y]_{GSEQ} > 0$. Both plots show the behaviour expected of the R-M effect on solar-wind magnetosphere coupling (dominated by magnetopause reconnection) with $[B_Y]_{GSEQ} < 0$, giving enhanced $P_\alpha/P_o$ at the March equinox ($F \approx 0.22$) and $UT$ of about 22hrs, whereas $[B_Y]_{GSEQ} > 0$ gives enhanced $P_\alpha/P_o$ at the September equinox ($F \approx 0.73$) and $UT$ of about 10hrs. The CXFORM Coordinate transformation package was also used to compute the GSEQ to GSM transformation of unit IMF vectors in the $+Y$ and $-Y$ directions of GSEQ and to give the IMF clock angle $\theta$ and hence the Russell-McPherron predictions of the $sin^4(\theta/2)$ IMF orientation factor in $P_\alpha$ and hence the factor $P_{RM}(F, UT)$. Note that in the original paper, *Russell and McPherron* (1973) used a half-wave rectified southward component IMF orientation factor ($B_S/B$) whereas we employ $A_\theta = sin^4(\theta/2)$: these two have been compared and discussed by *Lockwood et al.* (2020b). The black lines in the top row of Figure 16 are the contours of $1 + sin^4(\theta/2)$ of 1.28 and 1.31 and it can be seen that agreement is very close. However, the colour pixels in Figure 16a and 16b highlight an important point made by *Lockwood et al.* (2020b) namely that while the "favoured" equinox/ $UT$ shows a marked enhancement in $P_\alpha/P_o$ for a given polarity of $[B_Y]_{GSEQ}$ the "unfavoured" equinox/$UT$ shows a decrease in $P_\alpha/P_o$ that is almost as large. This is true for both polarities of $[B_Y]_{GSEQ}$ and *Lockwood et al.* (2020b) point out that the same thing is true – but, crucially, much less so for geomagnetic indices. Hence rather than it being the enhancement for the favoured equinox/$UT$ being the cause of the semi-annual variation (the traditional view of the R-M effect), it is the lack of a corresponding decrease for the unfavoured equinox/$UT$ that really generates the semi-annual variation. It is this fact that makes the semi-annual variation in geomagnetic indices a much larger fractional amplitude than that in $P_\alpha/P_o$. The effect of this is seen when we look at the R-M pattern in all data in Figure 16c. The R-M pattern is still seen in the data for both $[B_Y]_{GSEQ}$ polarities but it is not nearly as clear-cut and is made noisy by individual events of large negative $[B_Z]_{GSEQ}$, as pointed out by *Lockwood et al.* (2020a; b). Furthermore, the amplitude of the pattern is much reduced compared to that in the cases for the two $[B_Y]_{GSEQ}$ polarities separately: in Figures 16a and 16b, the amplitude of the $P_\alpha/P_o$ pattern is close to ±50%, whereas in Figure 16c is close to ±5%. This is why R-M effect is so much clearer when we sort data according to the $[B_Y]_{GSEQ}$ IMF component polarity than when we consider all data. This was one of the many valid points made by *Berthelier* (1976, 1990). We here scale the theoretical pattern of R-M forcing, $P_{RM}(F, UT)$ based on the IMF orientation factor $sin^4(\theta/2)$ (shown by the contour lines in Figure 8c) to match the amplitude of the pattern in the observed $P_\alpha/P_o$ which yields

$$P_{RM}(F, UT) = c_{RM} \, sin^4(\theta/2)/< c_{RM} \, sin^4(\theta/2) > \qquad (12)$$

and $c_{RM} = 0.1613$ gives the best least-square fit to the observations and the contours shown in black in Figure 16c for 1.045 and 1.050. The normalisation is needed to ensure the mean value of $P_{RM}(F, UT)$ over all $F$ and $UT$ is unity.

### 3-iii. *The dipole tilt and solar wind dynamic pressure factor, $P_\psi$*

Paper 2 (*Lockwood et al.*, 2020b) showed that the equinoctial pattern arises in the fit residuals of normalised power input to the magnetosphere, $P_\alpha/P_o$, to the $am$ index. This is not surprising, given that the equinoctial pattern is present in the $am$ data but not present in the $P_\alpha/P_o$ estimates. However, what is significant is that Paper 2 showed the amplitude of that equinoctial pattern in the fit residuals increases linearly with the solar wind dynamic pressure, $p_{sw}$. The effect of dynamic pressure was also clearly shown to be an independent effect to that of $P_\alpha$ because $am$ increases with $p_{sw}$ at all fixed values of $P_\alpha$. *Finch et al.* (2008) show the equinoctial pattern in data from high-latitude and auroral magnetometers arises on the nightside, increases with $p_{sw}$ and is associated with the auroral electrojet and substorm current wedge. In addition, *Chambodut et al.* (2013) showed that the amplitude of the equinoctial pattern in data from mid-latitude magnetometers is greatest at midnight and smallest at noon. Together these factors strongly indicate that geomagnetic activity is enhanced by increased $p_{sw}$ squeezing the near-Earth tail, as discussed by *Lockwood* (2013) and consistent with the effect of $p_{sw}$ on the lobe field and energy content of the near-Earth tail found by *Caan et al.* (1973) and *Karlsson et al.* (2000). The modelling presented in Paper 3 of the series supports this idea and shows that the effectiveness of the squeeze depends on the dipole tilt, $\psi$.

The factor $P_\psi(\psi)$ was modelled in Paper 3 (*Lockwood et al.*, 2020c) using the asymmetric magnetopause location model of *Lin et al.* (2017) by assuming that the tail is in equilibrium with a solar wind of dynamic pressure $p_{sw}$. These authors modelled the magnetic shear $\Delta B$ across the cross tail current sheet for various values of the geoeffective southward component of the IMF $[B_Z]_{GSM}$, and the solar wind dynamic pressure, $p_{SW}$ that are the inputs to the magnetopause model. The shape of the variation of $\Delta B$ with $\psi$ shown in Figure 6c of *Lockwood et al.* (2020c) does not vary with $[B_Z]_{GSM}$ nor $p_{SW}$ and a 4$^{th}$ order polynomial that fits (with appropriate scaling) all the variations of $\Delta B$ that is accurate to within an r.m.s. error of 0.02% is

$$P_{\psi\Delta B}(\psi) = 3.645 \times 10^{-8}\psi^4 - 1.059 \times 10^{-7}\psi^3 - 1.115 \times 10^{-4}\psi^2 + 4.768 \times 10^{-4}\psi + 1 \quad (13)$$

which is plotted as a function of $F$ and $UT$ in Figure 16d. *Lockwood et al.* (2020b) show that the equinoctial pattern in the $am$ index increases linearly in amplitude with solar wind dynamic pressure we fit the above functional form $P_{\psi\Delta B}(\psi)$ to the pattern amplitude the find for the conductivity-corrected $am$ observations ($am_{cc}$) for the mode value of $p_{SW}$ of 1.50 nPa using

$$P_\psi(\psi) = \{P_{\psi\Delta B}(\psi) - c_\psi\}/< \{P_{\psi\Delta B}(\psi) - c_\psi\} > \qquad (14)$$

where the mean value is over all $F$ and $UT$. The value of $c_\psi$ is the same for the northern and southern hemisphere sub-indices because $P_{\psi\Delta B}(\psi)$ is a measure of the whole tail and not just one tail lobe. Using the known pattern of $\psi(F, UT)$ this yield $P_\psi(F, UT)$ which is, as required, normalised to be unity. Note that increased $c_\psi$ means that the amplitude of the normalised $P_\psi(F, UT)$ pattern is increased.

### 3-iv. *The north and south pole motion factors, $[P_{PM}(F, UT)]_N$ and $[P_{PM}(F, UT)]_S$*

The sunward motion (in the $+X$ direction) in the GSEQ frame of the axial poles at speed $[V_X]_{HP}$ (where H is N for the northern hemisphere and S for the southern) generates a modulation to the transpolar voltage in that hemisphere (in the GSEQ frame):

$$\Delta\Phi_{cfH} = c_{PM}[\Delta\Phi_{PM}]_H = 0.9c_{PM}.d. < B_{iYZ} > .[V_X]_{HP} \qquad (16)$$

where $d$ is the polar cap diameter and $< B_{iYZ} >$ is the ionospheric magnetic field normal to the $X$ direction. The factor 0.9 allows for the fact that there is some dipole tilt motion in a geomagnetic frame, as discussed in Section (1-ii). The modulation of the transpolar voltage is by a factor $(1+ \Delta\Phi_{cfH}/ \Phi_{cf})$. The factor needed (that averages unity over all $F$ and $UT$) is

$$[P_{PM}(F, UT)]_H = 1+ \Delta\Phi_{cfH}/ \Phi_{cf} = 1+ (0.9c_{PM}.d. < B_{iYZ} > .[V_X]_{HP})/ \Phi_{cf} \qquad (16)$$

where we compute $\Phi_{cf}$ from the average $am$ of the interval in question from the regression equation given by equation (A4) in Appendix A of *Lockwood et al.* (2020b).

$$\Phi_{cf} = (6.68 \times 10^{-5})am^3 - (1.66 \times 10^{-2})am^2 + 1.89am + 6.17 \qquad (17)$$

Unlike for the constants $c_{RM}$ and $c_\psi$ we have no a-priori way to compute $c_{PM}$ and it is derived by matching the modelled $F$-$UT$ pattern to the (conductivity-corrected) observed $F$-$UT$ pattern.

## 4. Analysis

Equations (3), (12), (14), (15), (16) and (17) provide a full recipe for computing a model $F$-$UT$ pattern for the conductivity-corrected indices in the two hemispheres (generically $ax_{Hm}$) and the corresponding global index is the average of the two, $ax_m = (ax_{Nm} + ax_{Sm})/2$. We set the one remaining un-quantified and free variable, $c_{MP}$ in equation (15) by minimising the r.m.s. fit residuals for each index $\{< (ax_{Hm} - ax_{Hcc})^2 >\}^{1/2}$ for the 864 $F$-$UT$ averaging bins using the Nelder-Mead search method. In section 4-i we compare the model predictions with the $am$, $an$ and $as$ indices and in section 4-ii, the same procedure is used to model all 8 hemispheric $a\sigma$ indices (and hence the 4 global $a\sigma$ indices). The best-fit constants $c_\psi$ and $c_{PM}$ required for each index are given in Table 1. For both $an$ and $as$ (indeed all the $a\sigma$ indices, see next section) $c_{PM}$ is of order 0.1-0.2 which implies roughly 10-20% of the induced voltage caused by pole motion goes in to directly-driven changes in the transpolar voltage and 80-90% goes into the rate of change of field accumulation/loss in the near-Earth tail lobe. In Section 4-iii we study the effect of IMF $B_Y$ component and in Section 4-iv we study the relationship of average values to the occurrence of large events.

### 4.1 Geomagnetic response by hemisphere: Results for the *am*, *an* and *as* indices

The results from the model are shown in Figure 17, the right-hand column of which gives the $F$-$UT$ pattern for the conductivity-corrected observed index and the left-hand column gives the corresponding model prediction. The top row is for (b) the conductivity-corrected $an$ index, $an_{cc}$, and (a) its modelled equivalent $an_m$. The middle row is correspondingly for $as_{cc}$ and $as_m$ and the bottom row is for $am_{cc}$ and $am_m$. Figure 17 shows that the modelled $F$-$UT$ patterns match the observed ones very closely. The largest disagreement is for $an$, for which the predicted peak in the $UT$ variation at both equinoxes is near 13 hrs whereas in $an_{cc}$ it is near 16 hrs. This probably arises from the closeness of the northern geomagnetic dip pole from the rotation pole in recent decades (*Thébault, et al.,* 2015) which makes it likely that the eccentric dipole fit to the field is underestimating how far the longitude separation of the two poles has fallen from 180°.

Figure 18 shows the nett $UT$ variations obtained by integrating over all times of year, $F$. In each panel the colour scheme is red/blue/black lines are for the northern/southern/global index. Figure 18a shows the variations for the raw observed indices, $an$, $as$, and $am$ whereas Figure 18b is for the conductivity-corrected indices $an_{cc}$, $as_{cc}$, and $am_{cc}$. Comparison of Figures 18a and 18b shows the advantage of generating and modelling the conductivity-corrected indices. The variations in Figure 18a are quite complex but those in 18b are considerably simpler and we infer that the conductivity effects considerably complicate other hemisphere- and $UT$-dependent effects in the raw indices. The variation shown by the dot dash lines in Figures 18c and 18d are the variations for the modelled indices $an_m$, $as_m$, and $am_m$ and in Figure 18d are plotted on the same axes as $an_{cc}$, $as_{cc}$, and $am_{cc}$ to aid comparison. The agreement is very good, and the model is capturing the major features of the $UT$ variations of all three indices and even the difference in the peak times for $an_{cc}$ and $an_m$ is not as great as in Figure 17 after we have averaged over all $F$.

**4-ii.** *Analysis of geomagnetic response by MLT sector: Results for the $a\sigma$ indices*

The $a\sigma$ indices give us an opportunity to study the geomagnetic response in different MLT sectors in the same way. These indices are available for 1959-2013. The 4 global indices, $a\sigma(dawn)$, $a\sigma(noon)$, $a\sigma(dusk)$, and $a\sigma(midn)$, are available from the International Service of Geomagnetic Indices (ISGI) website, the 8 hemispheric sub-indices ($a\sigma_N(dawn)$, $a\sigma_S(dawn)$, $a\sigma_N(noon)$ … etc.) are available from ISGI on request. Figure 19 shows the $F$-$UT$ patterns for all 12 of the conductivity-corrected $a\sigma$ indices (hemispheric and global), corrected using the $F$-$UT$ conductivity factor patterns shown in Figure 15. Figure 20 shows the corresponding modelled variations. To model the $a\sigma$ indices we use the same procedure as used for the $am$, $an$ and $as$ indices described above. The best-fit constants $c_\psi$ and $c_{PM}$ are given in Table 1. In interpreting the various values of $c_\psi$ and $c_{PM}$ we should remember that the definitions used mean that larger values of $c_\psi$ and $c_{PM}$ increase the amplitudes of the $P_\psi(F, UT)$ and $P_{PM}(F, UT)$ patterns, respectively. For $c_\psi$ the only significant trend is that it is lower for $a\sigma$ noon, which is expected given that the equinoctial pattern arises on the nightside. The $c_{PM}$ values for noon are also slightly larger which means a slightly larger fraction of the voltage induced by poleward motion is appearing as a directly-driven ionospheric voltage. Values for the southern hemisphere are systematically a little larger than for the northern which may well be another symptom that the eccentric dipole is underestimating the north-south asymmetry. These constants are then used to model $a\sigma_N(x)$ and $a\sigma_S(x)$ in the same way that they were used to model $an$ and $as$ in the previous section. The modelled $a\sigma(x)$ is then the average of the two.

As for the $am$ indices, the agreement is good in all cases. Figure 21 shows the variations of the modelled (dot-dash lines) and conductivity-corrected $UT$ variations (solid lines) in the same format as Figure 17d. Parts (a), (b), (c) and (d) of Figure 21 are for the dawn, noon, dusk and midnight $a\sigma$ indices, respectively, and in all panels red, blue and black lines are for the north, south and global images. In general, the southern hemisphere indices are modelled more closely than the northern, but agreement is good in all cases.

### 4-iii. *Analysis of geomagnetic response by IMF Y-component polarity.*

We have also used the model to also simulate the results for the $[B_Y]_{\text{GSEQ}} > 0$ and $[B_Y]_{\text{GSEQ}} < 0$ data subsets. This further subdivision means there are 30 IMF polarity-index permutations and we here show the result for just the one case, the $a\sigma(midn)$ index, in Figure 22. The left-hand column is for IMF $[B_Y]_{GSEQ} < 0$, the middle column for $[B_Y]_{GSEQ} > 0$ and the right-hand column for all data. The top row shows the R-M solar wind forcing pattern $P_{RM}(F, UT)$ in each case. The middle panel shows the model predictions of the $a\sigma(midn)$ index and the bottom panel the corresponding averages of the $a\sigma(midn)$. It can be seen that agreement is good, and the model successfully predicts both sub-divisions of the data.

### 4-iv. *Large events and the effect of average levels of activity.*

*Lockwood et al.* (2019b) showed that the $F\text{-}UT$ patterns for average $am$ were matched by corresponding patterns in the occurrence of large events. Specifically, they studied the occurrence of events exceeding the $90\%$, $95\%$ and $99\%$ quantiles (q(0.9), q(0.95) and q(0.99), respectively). The data are continuously available for 1959-2019, inclusive (61 years) which is a total of 178250 3-hourly data samples and 534720 interpolated hourly samples. Each mean value in the 864 $F\text{-}UT$ bins used are therefore based on 618.9 samples per bin on average. Therefore, there are 61.89 samples above the $90\%$ quantile and just 6.19 samples in each bin above the $99\%$ quantile and hence derived $F\text{-}UT$ patterns become noisy because sample numbers are low, even for these data covering 61 years. In this section, we integrate over all $F$ and studying the average nett variation with $UT$ and the increases the above sample numbers by a factor of 36.

In the above analysis we used the overall mean of $am$ for the interval 1959-2019 to estimate the overall mean of the transpolar voltage, $< \Phi_{cf} >$ and so evaluate the fractional perturbation $\Delta\Phi_{cf}/< \Phi_{cf} >$. The above sample numbers for the $UT$ variation in $am$ are sufficient to allow us to divide the $am$ dataset into three 20-year intervals. To investigate the role of $< \Phi_{cf} >$, we here break the $am$ data into three 20-year intervals, 1960-1979, 1980-1999 and 2000-2019 for which average $am$ values are 21, 25 and 17 nT (by Equation (17), corresponding to $\Phi_{cf} \sim 39$, 43 and 34 kV) and we again compute $[V_X]_{NP}$ and $[V_X]_{SP}$ for the eccentric dipole axial pole locations at the middle of each interval.

The observed (black) and modelled (mauve) $UT$ variations (averaged over all $F$) are shown in parts d-f of Figure 23. These have not been conductivity-corrected because averaged of all $F$ and for the global $am$ index these corrections are negligibly small. The higher activity levels for 1980-1999 mean that $\Delta\Phi_{cf}/< \Phi_{cf} >$ is smaller and the pole motion terms $[P_{PM}(F, UT)]_N$ and $[P_{PM}(F, UT)]_S$ are less important and the $UT$ variation is dominated by the other two factors, $P_{RM}(F, UT)$ and, in particular, $P_{\psi}(F, UT)$. For 2000-2019 the lower activity level means that the pole motion terms have a much greater effect. For 1960-1979 the effect is halfway between that for the other two intervals. Parts a-c of Figure 23 compare the $UT$ variations in average fields (reproduced as dashed black lines) with the occurrence of large geomagnetic storms by giving the 90% quantile of the distribution of 3-hourly $am$ (q(0.9), cyan lines) and the 99.99% quantile (q(0.9999), blue lines). The events that meet the $am > $ q(0.9) criteria last for a total of 17532 hours in total in the 20-year intervals, whereas those that meet the $am > $ q(0.9999) criteria last for a total of just 17.532 hours. It can be seen that, despite their extreme rarity, even the latter are showing $UT$ variation consistent with the effect of pole motions that have been introduced here for the first time.

## 5. Discussion and Conclusions

We have identified for the first time a factor that introduces a systematic Universal Time variation into global geomagnetic activity but has been previously overlooked, namely the electric fields induced voltage Given that the largest disturbances occur in the midnight sector of Magnetic Local Time (MLT), this *UT* variation means that in geographic coordinates there is a longitudinal variation in geomagnetic activity which may be a useful fact in forecasting space weather and quantifying space weather risks. This also offers a potential explanation of some aspects of long-standing reports of longitudinal structure in auroral phenomena (e.g., *Berkey*, 1973; *Stenbaek-Nielsen*, 1974; *Luan et al.*, 2011; *Liou et al.*, 2018) although other aspects may be due to longitudinal structure in the field, such as the south-Atlantic anomaly.

If Earth's field were a symmetric, Earth-centred dipole, the nett effect would be relatively straightforward. The two polar caps and tail lobes would undergo matching diurnal cycles in which the pole moved against, and then with, the solar wind flow and these cycles would be in antiphase in the two hemispheres. In the first half of this cycle, the directly-driven component would mean the voltage appearing across one polar cap would be increased while that in the other would be correspondingly decreased, a situation that would be symmetrically reversed in the second half of the cycle. The storage/release response would mean that the energy being stored in the tail lobe of one hemisphere would also be increased and then decreased whilst the variation for the other lobe would be of equal magnitude and in antiphase. Hence for this effect in isolation, the average transpolar voltage for the two poles would be constant, as would the total rate of storage of tail energy, as while the northern hemisphere lobe was gaining additional stored energy the southern would be losing it (or at least gaining it more slowly) and vice versa. Hence the effect would be to alternately pump additional energy into one tail lobe and then the other, in each case recovering it during the other half of the cycle.

However, Earth's magnetic field is increasingly unlike a symmetric, Earth-centred dipole and we have here used an eccentric dipole model to investigate the effects of the (growing) hemispheric asymmetry. Because the offset from the rotational pole of the southern magnetic pole is roughly double that for the northern pole, these effects are roughly twice as large in the southern hemisphere than in the northern. Because both follow regular diurnal cycles the effects are averaged out over a full day, but they do not cancel on timescales below a day, leaving a strong *UT* variation. This asymmetric loading and unloading of one lobe relative to the other is likely to be associated with short-lived asymmetries in auroral precipitation (*Laundal and Østgaard*, 2009, *Laundal et al.*, 2010) given that asymmetric lobe flux content caused by a strong *Y*-component of the IMF (*Cowley et al*, 1991) has been shown to be the cause of non-conjugate auroral behaviour (*Reistad et al.*, 2013).

The new proposed mechanism is different to longitudinal variations in auroral ionospheric dynamics associated with hemispheric asymmetries in local magnetic field strength and direction (*Gasda and Richmond*, 1998) or due to ionospheric conductivities (*Lyatsky et al.*, 2001; *Newell et al.*, 2002). In the present paper, we do not attempt to make detailed comparisons of the mechanism we propose with other proposals; however, we do note that the possibility exists that various phenomena that have in the past been attributed to ionospheric conductivity effects and/or ionosphere-thermosphere momentum exchange and/or longitudinal structure in the geomagnetic field may have been associated with the effects of polar cap motions that were not considered.

We have compared our first-order model of the combined effects with observed mid-latitude range indices, using the $an$, $as$, and $am$ indices that are available for 61 years now. Using 36 equal-sized bins of fraction of calendar years, $F$, this means we have 619 interpolated 1-hour samples in each $F$-$UT$ averaged bin. The twelve $a\sigma$ indices are only available for 1959-2013 (55 years) and so this number drops to 558 samples in each bin. These large sample numbers are important. *Lockwood et al.* (2020a; 2020b) show that the largest geomagnetic events are not caused by the R-M mechanism but rather by events of strong southward IMF in the GSEQ frame: indeed, these authors show that the R-M effect actually reduces the geoeffectiveness for the most southward-pointing fields in GSEQ. These large events are mainly driven by field inside, and ahead of, coronal mass ejections and the impact of such an event on Earth must be completely random in $UT$ and there is no evidence that the expectation that it is also random in $F$ is incorrect. The large sample numbers in each bin are important to average out the random occurrence of the most geo-effective solar wind hitting Earth.

We have demonstrated how the four factors discussed in this paper can explain why average geomagnetic activity displays equinoctial (McIntosh) time-of-year/time-of-day patterns, with an additional $UT$ variation, instead of the Russell-McPherron pattern. This is despite the fact that sorting the data by the *Y*-component of the IMF reveals the Russell-McPherron effect is the fundamental cause of the semi-annual variation, and of the $UT$ variation for each equinox separately, as demonstrated in Papers 1 and 2 (*Lockwood et al.*, 2020a; b). The factors included in our initial modelling of the patterns are the Russell-McPherron effect; ionospheric conductivity variations; the dependence of tail squeezing on dipole tilt angle $\psi$ and dynamic pressure the hemispheric asymmetry in the geomagnetic field. The last of these factors acts in two ways. The first is the effect discussed in Paper 3 on the tail squeeze, but the new effect introduced here is that it also generates a hemispheric asymmetry in the diurnal cycles of sunward motion of the two poles.

Given that the incidence of the most geoeffective solar wind impacting on the magnetosphere should be random in both $F$ and $UT$, and that the Russell-McPherron effect (that could introduce such variations) has little, or even the inverse, effect when the field is most strongly southward in the GSEQ frame (*Lockwood et al.*, 2020a;b), the variations in the occurrence of the largest storms with both $F$ and $UT$ is a puzzle. *Cooker et al.* (1992) proposed a solution to the dichotomy of the R-M effect giving the semi-annual variation, and yet that large storms are driven by CME impacts, by proposing that the IMF $[B_Y]_{\text{GSEQ}}$ component is enhanced by compression in the sheath ahead of the impacting CME and this enhanced $[B_Y]_{\text{GSEQ}}$ is converted into enhanced $[B_S]_{\text{GSM}}$ at the favoured equinox. However, this is not the answer to the puzzle because we here show that enhanced $[B_S]_{\text{GSM}}$ is associated with enhanced negative $[B_Z]_{\text{GSEQ}}$ and there is no enhancement in either average values or events of large negative $[B_Z]_{\text{GSEQ}}$ at the equinoxes, nor is there any proposed reason why there might be. This was discussed specifically in the context of the semi-annual variation of large storms in the previous papers in this series, but Figure 23 shows it is an issue in relation to the $UT$ variation as well. More work is needed to understand why a model that is aimed at predicting average geomagnetic activity levels is generating a $UT$ variation that even approximates to the $UT$ variation in the occurrence of the largest storms. There are a number of possibilities, including the effect of "pre-conditioning" of the magnetosphere-ionosphere system by average conditions ahead of the arrival of an event at Earth. A key part of the $UT$ variation modelled here is the sinusoidal sunward motion of the poles that is here introduced for the first time and hence we need to study how this mechanism influences large events as well as the average conditions considered in this paper.

Lastly, we can now look back with the benefit of hindsight at the debate in the literature about the R-M and equinoctial patterns. Annick Berthelier was correct in the main point of her comment (*Berthelier*, 1990) on the paper by *Russell* (1990) that there was no element of the latter paper that showed the equinoctial effect (what she called the "McIntosh effect") was not operating alongside the R-M effect. In that comment, and in her 1976 paper (*Berthelier*, 1976), she demonstrated that she fully understood the R-M effect was active because of the observed influence of the IMF $Y$-component. It is not clear why in their response *Russell and Scurry* (1990) were so adamant that the "McIntosh effect" was not a factor; however, one interesting point to note here is the influence of semantics on the debate and what was meant by the term "McIntosh effect". *Russell and Scurry* (1990) ascribe to the McIntosh effect an invocation of the Kelvin-Helmholtz (K-H) instability on the magnetopause. However, Berthelier makes no statements specifically invoking this mechanism and indeed the term never appears in the original paper by *McIntosh* (1959) who simply, and correctly, pointed out a dependence of geomagnetic activity on the dipole tilt angle. The K-H mechanism was introduced into the story 11 years after the McIntosh paper by *Boller and Stolov* (1970) and it is this that seems to be the main objection of Russell and Scurry, and the results of the present series of papers confirm that this is a fully valid objection. In her 1976 paper, Berthelier mentions that the K-H mechanism had been postulated but never attempts to quantify its effectiveness and never specifically invokes it. It is not mentioned in her 1990 comment at all. Hence, for example, by postulating dipole tilt effects on magnetopause reconnection, *Russell et al.* (2003) were adding precisely what Berthelier meant by the "McIntosh effect" to the R-M effect, i.e., a dependence on the dipole tilt angle. In fact, the results of *Finch et al.* (2008) subsequently showed that the McIntosh effect is a nightside phenomenon and does not influence dayside high-latitude flows and currents in the way that a dipole tilt effect on magnetopause reconnection would. *Finch et al.* (2008) and *Lockwood* (2013) ascribe the effect to the substorm current wedge in the near-Earth tail. This is fully consistent with Berthelier's assertion that the R-M and McIntosh/equinoctial effect could both be operating simultaneously.

In the present paper and in Paper 3 we have shown how the pressure equilibrium in the near-Earth tail for a hemispherically-asymmetric geomagnetic field explains how the energy input into the magnetosphere, controlled by the R-M mechanism, results in a "McIntosh pattern" and an additional UT variation. There is no need to invoke the K-H instability, nor any other "viscous-like" (meaning anything that is not reconnection) interaction across the magnetopause. Indeed, the results of *Finch et al.* (2008) indicate that the equinoctial/McIntosh pattern has nothing to do with solar wind-magnetosphere coupling at all. In this context, the Expanding-Contracting Polar Cap (ECPC) model also indicates that viscous-like interaction voltages have, in the past been greatly overestimated in studies that overlook the delayed response of the tail. *Lockwood et al* (1990) pointed out that the ECPC model predicts that ongoing reconnection in the tail contributes to transpolar voltage which can therefore have large values even after the IMF has turned northward. *Wygant et al.* (1983) showed (their Figure 6) that one hour after a northward turning of the IMF the transpolar voltage could range between 10kV and 100kV, but that the upper limit decayed with time elapsed since the northward turning, such that after about 10 hours little more than 10kV was observed. This was explained by *Lockwood et al.* as the transpolar voltage being enhanced in some cases by substorm expansion phases after the IMF had turned northward as energy stored in the tail is released, but such substorms became weaker as the interval of continuous northward IMF progressed because open flux lost was not replenished, a conclusion supported by the analysis of *Milan* (2004). *Lockwood* (2019) has pointed out that the geomagnetic tail never disappears and so there is always some magnetic shear between open field lines of the two tail lobes and so, almost certainly, ongoing tail reconnection at some level. This means that even much of the 10kV seen more than 10 hours after the IMF was last southward is likely to be reconnection-driven and not caused by any viscous-like interaction across the magnetopause. Hence the ECPC model shows that residual voltage seen when the IMF is northward cannot be used as evidence for a viscous-like transfer of momentum across the magnetopause. Similarly, the results presented in this paper show that a McIntosh/equinoctial pattern in geomagnetic activity cannot be used as evidence for a viscous-like interaction.

**Acknowledgements and data and software sources**. The authors are grateful to the staff of the International Service of Geomagnetic Indices (ISGI), France and collaborating institutes for the compilation and data basing of the $am$ and $a\sigma$ indices which were downloaded from http://isgi.unistra.fr/data_download.php and to the staff of the Space Physics Data Facility (SPDF) at NASA's Goddard Space Flight Center for the Omni composite of interplanetary observations (made available by SPDF from https://omniweb.gsfc.nasa.gov/ow_min.html ). For the SuperMAG indices data we gratefully acknowledge the PIs and staff of the many groups contributing data. SuperMAG data are available from http://supermag.jhuapl.edu/indices/?layers=SME.UL. The homogeneous *aa* index, $aa_H$, is available as 3-hourly value or daily means from http://www.personal.reading.ac.uk/~ym901336/pdfs/361_Lockwood2_SupplementaryMaterial_newaa_3hourly.txt . We are also grateful to the World Data Center for Geomagnetism, Kyoto for generating and making available the *AL* indices. (http://wdc.kugi.kyoto-u.ac.jp/). The SUBSOL routine of the LOWTRAN7 Sun and Moon Models package was coded for Matlab use by Meg Noah of the US Air Force Geophysics Laboratory in 2019 and is available from https://www.mathworks.com/matlabcentral/fileexchange/71203-lowtran7-sun-and-moon-models?s_tid=FX_rc1_behav. The CXFORM Coordinate transformation package originally written by Ed Santiago of Los Alamos National Laboratory and Ryan Boller of NASA's Goddard Space Flight Centre and re-programmed for Matlab by Patrik Forssén (SatStar Ltd & Karlstad University) in 2017, is available from https://spdf.sci.gsfc.nasa.gov/pub/software/old/selected_software_from_nssdc/coordinate_transform/#Mi.

 This work is supported by a number of grants. The work of ML, LAB and MJO at University of Reading is supported by the SWIGS NERC Directed Highlight Topic Grant number NE/P016928/1/. MJO, CJS and ML at the University of Reading are also supported by STFC consolidated grant number ST/M000885/1. Funding for KAW at University of Saskatchewan was provided by the Canadian Foundation for Innovation (CFI), the Province of Saskatchewan, and a Discovery Grant from the Natural Sciences and Engineering Research Council (NSERC) of Canada. Initial work by KAW for this paper was carried out at University of Reading on sabbatical leave from University of Saskatchewan. The work of OC at École et Observatoire des Sciences de la Terre (EOST) is supported by CNES, France. CH is supported on a NERC PhD studentship as part of the SCENARIO Doctoral Training Partnership.

**Table 1**. Best-fit coefficients used to derive modelled patterns of geomagnetic activity

| Index | $c_\psi$ | $c_{PM}$ for the northern polar cap | $c_{PM}$ for the southern polar cap |
|---|---|---|---|
| $am$ | 0.78 | 0.10 | 0.11 |
| $a\sigma(dawn)$ | 0.85 | 0.09 | 0.09 |
| $a\sigma(noon)$ | 0.50 | 0.29 | 0.33 |
| $a\sigma(dusk)$ | 0.80 | 0.22 | 0.28 |
| $a\sigma(midn)$ | 0.85 | 0.22 | 0.25 |

**Appendix A. Summary of the role of solar wind dynamic pressure in generating the semi-annual variation**

In this Appendix we present some alternative graphics to make the same points as were made using other plots in Papers 1, 2, and 3 (and brought together in the present paper) about how solar wind dynamic pressure contributes to the semi-annual variation. Figure 1 of Paper 3 (Lockwood et al., 2020c) shows that geomagnetic activity (both on average and in the occurrence of large events) is enhanced at constant power input to the magnetosphere (which depends on IMF orientation) by enhanced solar wind dynamic pressure. This is also supported by the modelling shown in Paper 3 which shows that, for a given magnetospheric open flux, energy stored in the tail and cross-tail current is increased by enhanced solar wind dynamic pressure, a result first reported from observations by *Caan et al.* (1973). Another plot which explains the relationship of the effects of solar wind dynamic pressure and of power input to the magnetosphere is Figure 19 of Paper 2 (*Lockwood et al.,* 2020b).

Figure A1 shows the occurrence distributions of two IMF parameters for a full Hale cycle of near-continuous IMF observations for the interval 1996-2018. Figure A1a is the distribution of values of the IMF $B_Y$ component in the GSEQ frame, $[B_Y]_{\mathrm{GSEQ}}$ and shows that negative values are very slightly more common than positive ones, the asymmetry being mainly around the peaks that are at small $|[B_Y]_{\mathrm{GSEQ}}|$.

Figure A1b shows the distribution of the IMF clock angle in GSEQ , $\theta_{\mathrm{GSEQ}} = tan^{-1}([B_Y]_{\mathrm{GSEQ}}/[B_Z]_{\mathrm{GSEQ}})$. The average value and the mode value of this distribution is 90° (the vertical red dashed line), which is the value assumed to apply all of the time in the demonstration of the R-M effect in the original paper (*Russell and McPherron*, 1973) . Figure A1b shows that, in reality, a full range of $\theta_{\mathrm{GSEQ}}$ values are seen with zero and 180° (respectively, purely northward and southward IMF in the GSEQ frame) being roughly half as common as $\theta_{\mathrm{GSEQ}} = 90°$. Appendix B of Paper 1 studied the implications of the fact that $\theta_{\mathrm{GSEQ}}$ is not always 90° (*Lockwood et al.,* 2020a).

Figure A2 makes the point that this roughly symmetric distribution of the two IMF $[B_Y]_{\mathrm{GSEQ}}$ polarities is present all the time during the interval covered. The data are means in half-year intervals centred on the two equinoxes. The top panel shows the sunspot number, the second panel the solar polar fields, the vertical grey lines marking the solar polar field polarity reversals. The third panel shows the fraction of the time, $f$, that the two $[B_Y]_{\mathrm{GSEQ}}$ polarities are present. It can be seen that $f$ is close to 0.5 for the two polarities all the time, there being largest deviations in the rising phases of the solar cycles. The senses of these deviations are different in the two solar cycles and reflects the solar polar field polarity, an indication of the Rosenberg-Coleman effect at work on $[B_X]_{\mathrm{GSEQ}}$ (along with the fact that $[B_Y]_{\mathrm{GSEQ}}$ and $[B_X]_{\mathrm{GSEQ}}$ most often have opposite polarity because of the Parker spiral configuration of the IMF). Note that the two solar cycles differ in amplitude and this will leave some net bias in the distribution for the whole Hale cycle, shown in Figure A1a.

The bottom panel of Figure A2, shows the number of 1-minute $[B_Y]_{\mathrm{GSEQ}}$ samples, $N$ (on a logarithmic scale), in the six-month intervals and in bins of $[B_Y]_{\mathrm{GSEQ}}$ that are 1nT wide. For large $|[B_Y]_{\mathrm{GSEQ}}|$ the distributions are highly symmetrical, but there are weak asymmetries at small $|[B_Y]_{\mathrm{GSEQ}}|$. The point relevant to the R-M effect is that for either equinox the distribution of $|[B_Y]_{\mathrm{GSEQ}}|$ values for the two polarities is broadly the same: this means that in order for there to be a semi-annual variation, with equinox peaks, the increase in geomagnetic activity for the favoured IMF $[B_Y]_{\mathrm{GSEQ}}$ polarity (for that equinox) must exceed the decrease in geomagnetic activity for the unfavoured $[B_Y]_{\mathrm{GSEQ}}$ polarity.

Figure A3 studies variations with fraction of a calendar year, $F$, for the two polarities of $[B_Y]_{\mathrm{GSEQ}}$. In every case, the plots have been normalised to the maximum value to help reveal the differences in the behaviour of the minimum values. The blue lines and symbols are for the *am* geomagnetic index and clearly show that the decrease for the unfavoured polarity at a given equinox is smaller in magnitude than the increase for the favoured polarity. It is this fact that gives the large semi-annual variation in *am*.

The orange lines and symbols give the variation predicted using the Russell-McPherron paradigm. It shows the half-wave rectified southward field in GSM, $[B_S]_{RM}$, computed by adopting the assumption that $[B_Z]_{GSEQ} = 0$ (so the clock angle in GSEQ, $[\theta]_{GSEQ} = 90°$), resulting in a northward field in GSM of $[B_Z]_{RM} = [B_Y]_{GSEQ} \sin(\beta_{GSEQ})$, where $\beta_{GSEQ}$ is the rotation angle between the GSEQ and GSM frames. This yields half-wave rectified southward field of $[B_s]_{RM} = -[B_Z]_{RM}$ for $[B_Z]_{RM} < 0$ and $[B_s]_{RM} = 0$ for $[B_Z]_{RM} \geq 0$. A baselevel value $b$ has been added to match the peaks of $am$ and of the other parameters for the favoured equinox at that polarity. This R-M prediction matches the observed variations for $am$ rather well.

However, if we do not make the simplifying assumption that $[\theta]_{GSEQ} = 90°$, and instead use the actual values shown in Figure A1b, we obtain the average variations for the half-wave rectified southward field in GSM, $[B_S]_{GSM}$, that are shown by the mauve lines and symbols. In this case, the decrease for the unfavourable polarity of $[B_Y]_{GSEQ}$ is almost equal in magnitude to the increase for the favourable $[B_Y]_{GSEQ}$ and when we add them together with the roughly matching probability distributions shown in Figure A2, we will get only a very small semiannual variation in the average value. Hence it is not just the half-wave rectification that is giving the good agreement to the semi-annual variation for the R-M effect demonstration, the simplifying assumption of $[\theta]_{GSEQ} = 90°$ is a vital component, and without it the semi-annual variation predicted is very small. The explanation of why this occurs for general $[\theta]_{GSEQ}$ was given in the Appendix B to Paper 1. (*Lockwood et al.,* 2020a).

The black lines and symbols in Figure A3 are for the estimated power input into the magnetosphere , $P_\alpha$. In this case the decrease for the unfavourable polarity of $[B_Y]_{GSEQ}$ is somewhat smaller in magnitude than the increase for the favourable $[B_Y]_{GSEQ}$, but not by as much as for the *am* index. Hence using $P_\alpha$ (with its $sin^4([\theta]_{GSEQ}/2)$ IMF orientation factor) is solving a small part of the anomaly introduced by using all values of $[\theta]_{GSEQ}$ (i.e. allowing for the non-zero IMF $[B_Z]_{GSEQ}$ component) and not assuming $[\theta]_{GSEQ} = 90°$, as in the original demonstration of the R-M effect.

Figure A4 shows how solar wind dynamic pressure $p_{SW}$ allows for most of the remaining anomaly. This plot has the same format as Figure A3, and the black lines and symbols reproduce the $P_\alpha$ variations shown in Figure A3. The red, green and blue lines show the variations of the averages of the $am$ index for simultaneous $p_{SW}$ in its three tercile ranges. The red line is for $p_{SW}$ in its lower tercile of values, the green line is for $p_{SW}$ in its middle tercile of values and the blue line is for $p_{SW}$ in its upper tercile. As in Figure A3, the peaks for the favoured equinox (for a given $[B_Y]_{GSEQ}$ polarity) are very similar indeed in shape (remember the variations have been normalised so that they all have the same magnitude), but the minima are very different. Figure A4 clearly shows that enhanced $p_{SW}$ increases values of $am$ during the unfavoured equinox and it is this that gives the larger semi-annual variation in $am$ than in $P_\alpha$ (and a very much larger semi-annual variation than that in $[B_S]_{GSM}$). Figure A4 is showing the effect on the variations through the year of the $p_{SW}$ effect that is noted in Figure 7 of the main paper.

This Appendix has used different graphical plots (of the same data) to those used before in this series of papers to emphasise the key point that the half-wave rectification of solar-wind-magnetosphere coupling is not the main cause of the semi-annual variation and that previous studies, which implied or stated that it was, were actually getting good agreement through the simplifying assumption that the IMF always lay in the solar equatorial frame. In reality, the key element introducing the asymmetry between the increased geomagnetic activity for the favourable IMF $B_Y$ component polarity and the unfavourable one is the solar wind dynamic pressure not a half-wave rectified coupling function. Solar wind dynamic pressure increases geomagnetic activity, for a given level of open flux production, at both equinoxes (favoured and unfavoured) - as is shown in Figure 7 of the main paper. Hence it increases the peak for the favourable IMF $B_Y$ component polarity whilst reducing the minimum for the unfavoured IMF $B_Y$ component polarity, causing the asymmetry required to generate the semi-annual variation. The modelling in Paper 3 (*Lockwood et al.*, 2020c) explains how the squeezing of the tail by solar wind dynamic pressure is less effective at the solstices than at the equinoxes because at the solstices a smaller fraction of the open flux present at any one time has migrated into the tail in the summer hemisphere (i.e., a larger fraction threads the dayside summer magnetopause). This is because, initially, after reconnection the sheath flow and curvature forces act in opposite directions on newly-opened field lines in the summer hemisphere, whereas they act in the same direction in the winter hemisphere. This difference and the solar wind dynamic pressure effect also explains the observed equinoctial (a.k.a. McIntosh) pattern in geomagnetic activity.

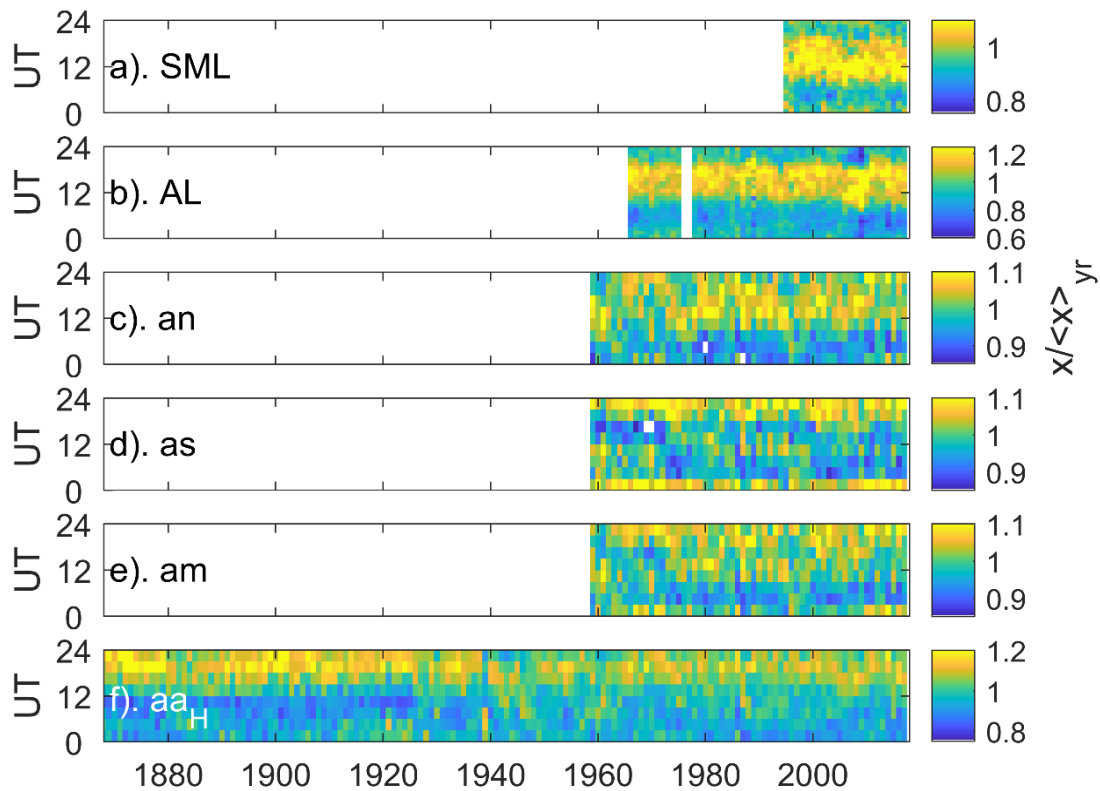

**Figure 1**. Universal Time-year spectrograms of normalised geomagnetic activity indices. In each panel the mean value in 3-hour bins of $UT$ for a given calendar year are shown as a ratio of the overall mean for that year (generically $x/<x>_{yr}$ where $x$ and $<x>_{yr}$ are, respectively, 3-hour and 1-year means of the index in question), colour-coded as a function of $UT$ and year. (a) the SuperMAG $SML$ index. (b) the auroral electrojet $AL$ index. (c) the northern hemisphere component of the $am$ index, $an$. (d) the southern hemisphere component of the $am$ index, $as$. (e) The $am$ index, $am = (an + as)/2$. (f) The homogenous aa index, $aa_H = (aa_{HN} + aa_{HS})/2$.

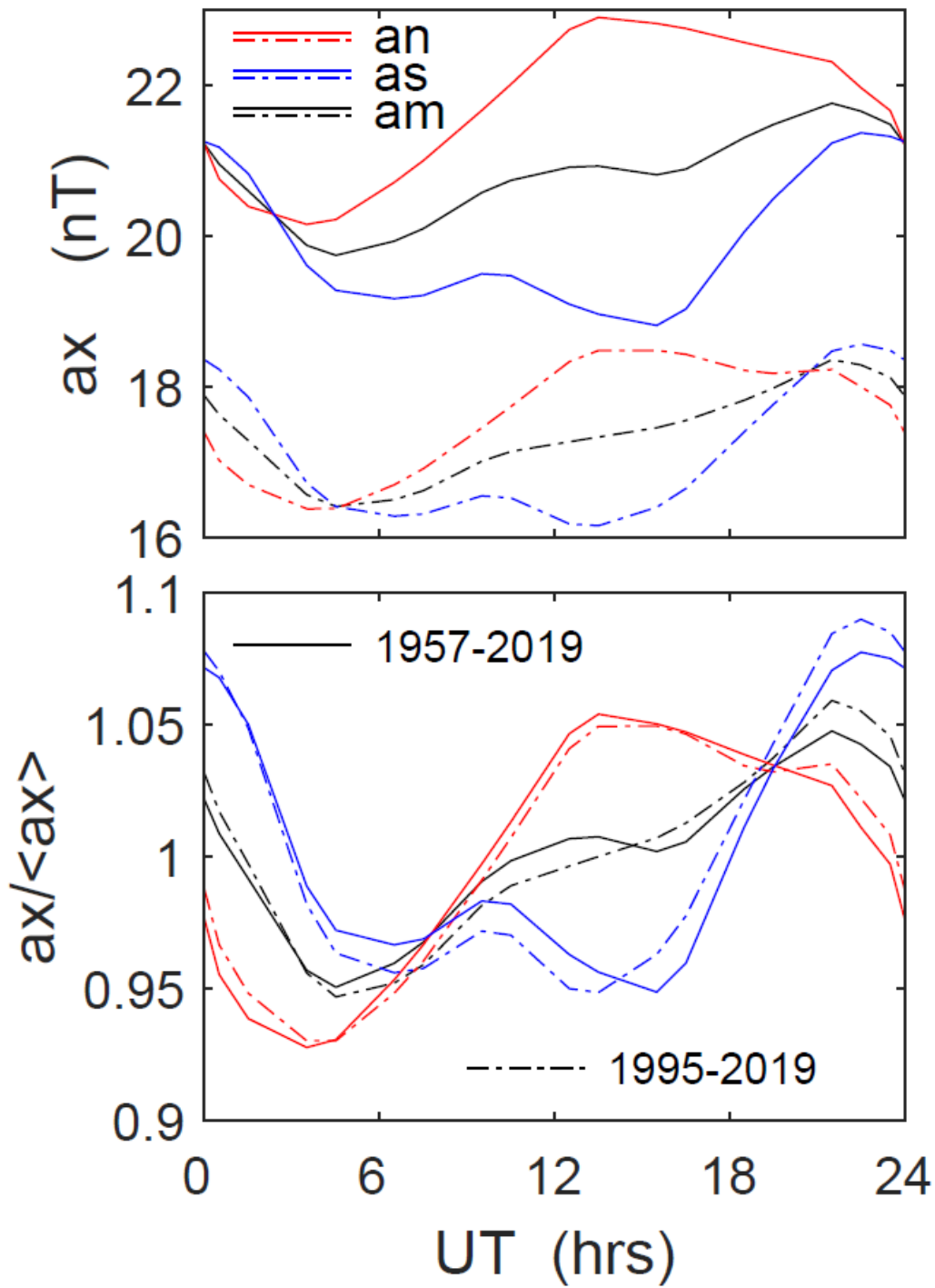

**Figure 2**. The *UT* variations of means of the (black lines) *am* index and its two hemispheric sub-indices *an* (for the northern hemisphere, red lines) and *as* (for the southern hemisphere, blue line). The solid lines are for 1959-2017, the dot-dash lines for 1995-2019. The top panel shows the mean values absolute values of the index (generically termed *ax*) in one-hour bins. The lower panel shows the mean values as a fraction of the overall mean for the interval, $ax/<ax>_{all}$  The hourly values were obtained by linearly interpolating the three-hour index values.

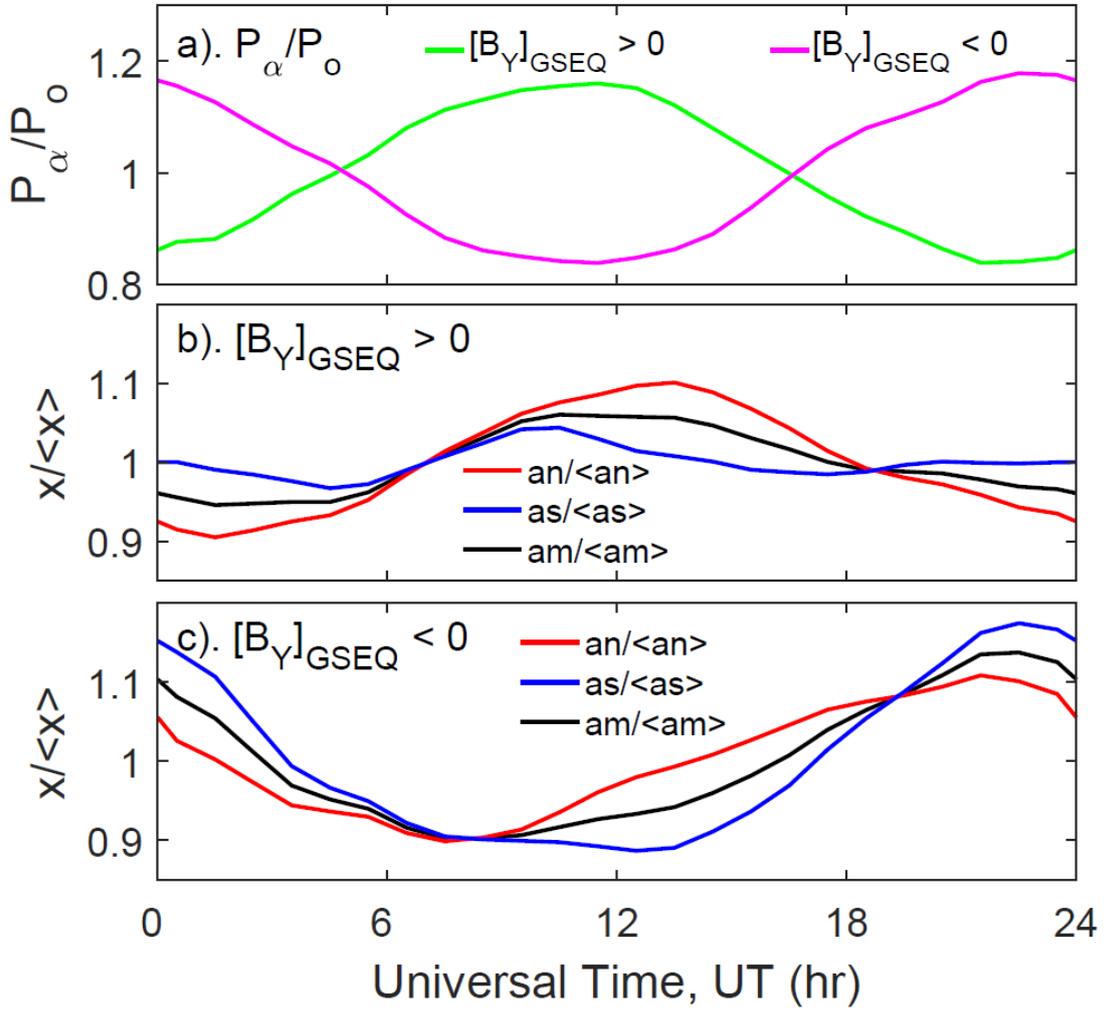

**Figure 3**. Universal Time variations sorted by the prevailing polarity of the IMF in the GSEQ frame ($[B_Y]_{GSEQ}$, averaged over the prior hour). (a) The normalised estimated power input into the magnetosphere, $P_\alpha/P_o$, for (green line) $[B_Y]_{GSEQ} > 0$, and (mauve line) $[B_Y]_{GSEQ} < 0$. (b) The fractional variation of the geomagnetic indices for $[B_Y]_{GSEQ} > 0$. (c) The fractional variation of the geomagnetic indices for $[B_Y]_{GSEQ} < 0$. Black lines are for the *am* index, the red lines for the *an* index and the blue lines are for *as*.

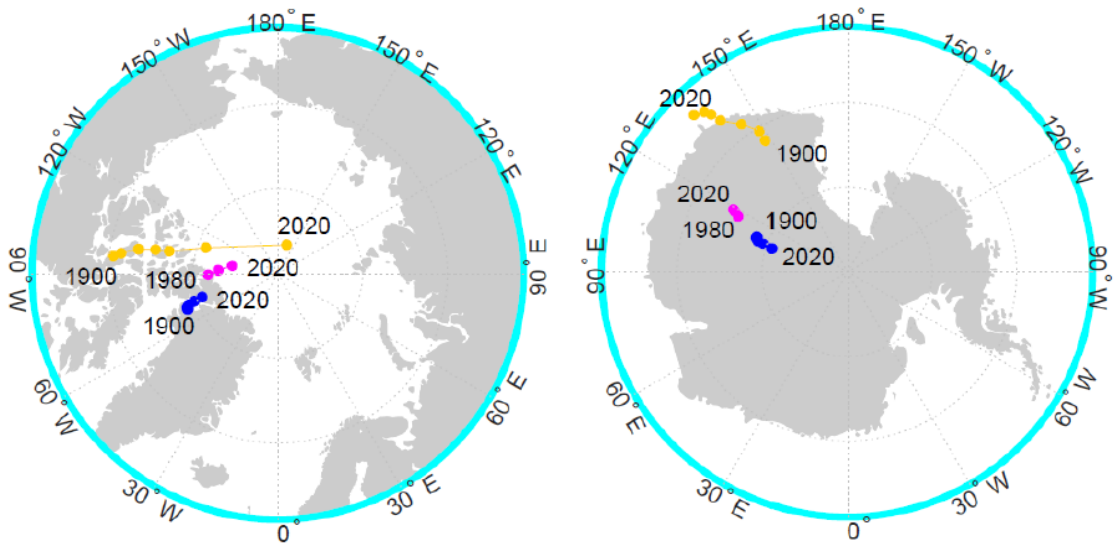

**Figure 4**. Maps of the geographic locations of (left) northern and (right) southern magnetic poles for various years. The orange and blue points are dip and geocentric dipole pole locations from the 12th generation of the International Geomagnetic Reference Field (IGRF) for 1900 to 2020 in steps of 20 years (from *Thébault et al.*, 2015). The mauve points are the axial poles for the years 1980, 2000 and 2020 from the eccentric dipole model fits (for which the dipole axis is not constrained to pass through the centre of the Earth) of *Koochak and Fraser-Smith* (2017)

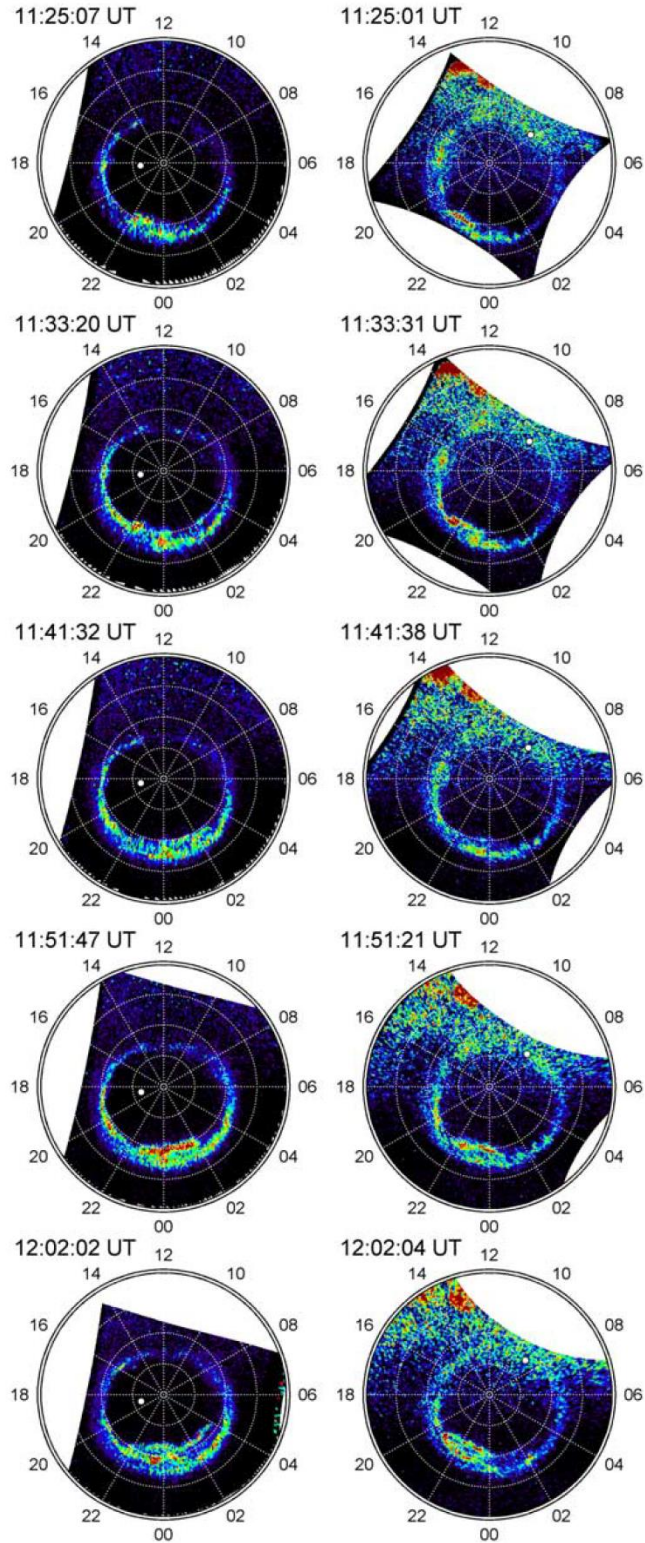

**Figure 5.** Two series of near-simultaneous images of the two auroral ovals observed between 11:24 and 12:10 UT on 23 October 2002. The left column shows observations of the northern hemisphere oval made by the FUV-SI13 instrument on the IMAGE satellite and the right column shows the series of near-simultaneous images of the southern hemisphere auroral oval made by the VIS-EC, instrument on the Polar satellite. respectively. Each image is shown in the geomagnetic latitude-magnetic local time (MLT) frame (using AACGM coordinate system) and white dot gives the location of the geographic pole. (from *Stubbs et al.*, 2005)

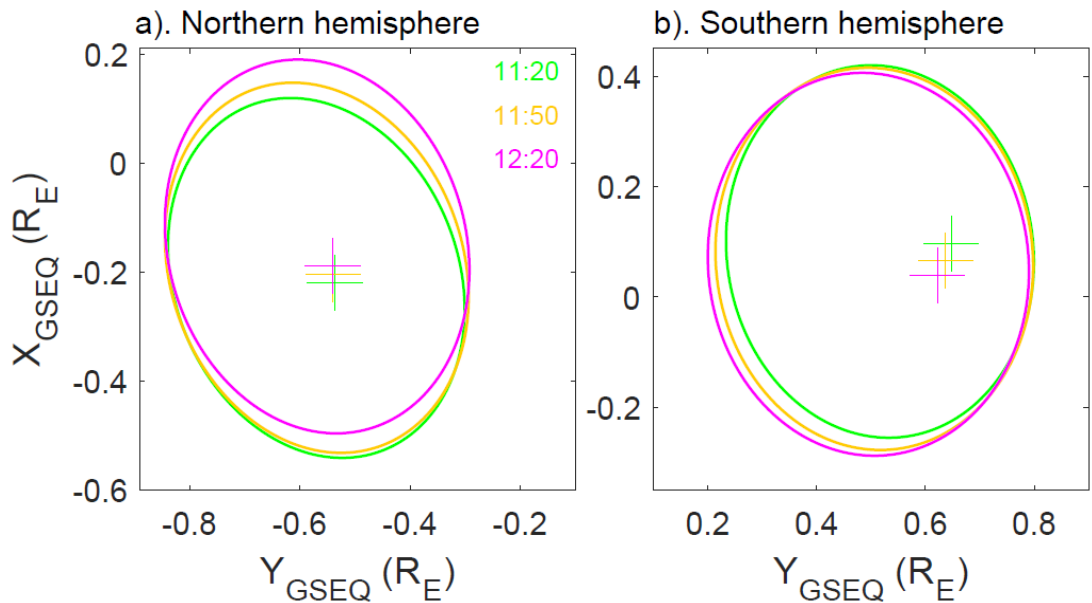

**Figure 6**. The fitted circular poleward edges of the aurora in the AACGM geomagnetic latitude-MLT frame at three times, half an hour apart (11:20 in green, 11:50 in orange and 12:20 in mauve) on 23 October 2002 (as fitted by *Stubbs et al.*, 2005) transformed into the GSEQ *XY* frame (where $X_{GSEQ}$ points towards the Sun). The crosses show the location of the eccentric dipole axial pole mapped in the same way and displayed using the corresponding colour. The mapping is for an assumed emission altitude of 130 km: (a) is for the northern hemisphere oval and shows both the pole and the oval moving toward the Sun; (b) is for the southern hemisphere oval and shows both the pole and the oval moving away from the Sun and in the and in the $-Y_{GSEQ}$ direction. Note that the $Z_{GSEQ}$ direction is into the plane of the diagram and so the *XY* plane is here viewed from the southern side and so dawn is the to the left and dusk to the right and noon to the top, (in the $+X$ direction) and hence the motion of the poles with *UT* is clockwise.

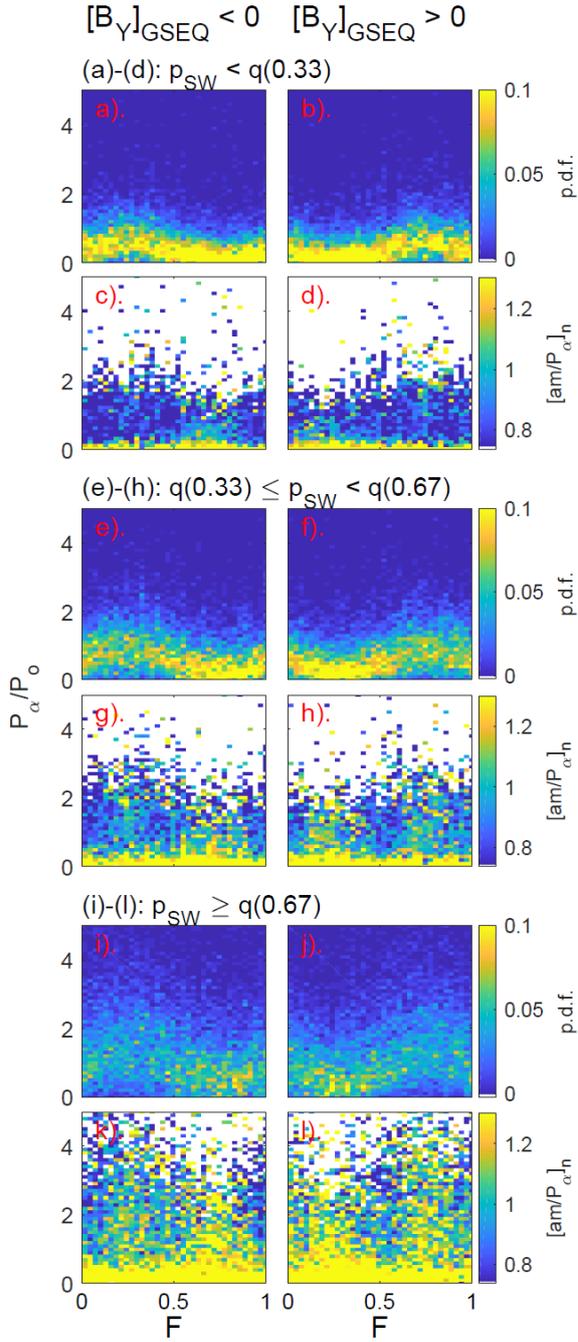

**Figure 7**. Parts a, b, e, f, i and j are plots of probability distribution functions of $(P_\alpha/P_o)$ as a function of $F$ and beneath each (parts c, d, g, h, k and l) is the corresponding plot of the normalized *am* amplification factor, $[am/P_\alpha]_n = (am/<am>_{all})/(P_\alpha/P_o)$, as a function of $F$ and in the same $(P_\alpha/P_o)$ bins as the p.d.f.s. The left hand panels are for IMF $[B_Y]_{GSEQ} < 0$, the right hand panels are for IMF $[B_Y]_{GSEQ} > 0$. The plots are in 3 groups of 4: parts a-d are for the lower tercile of the simultaneous (allowing for the propagation lag) solar wind dynamic pressure, $p_{SW} < q(0.33)$ ; parts e-h are for the middle tercile of the simultaneous solar wind dynamic pressure, $q(0.33) \leq p_{SW} < q(0.67)$; parts i-l are for the upper tercile of the simultaneous solar wind dynamic pressure, $p_{SW} \geq q(0.67)$, where $q(x)$ is the $x^{th}$ quantile of the overall distribution of $p_{SW}$ values.

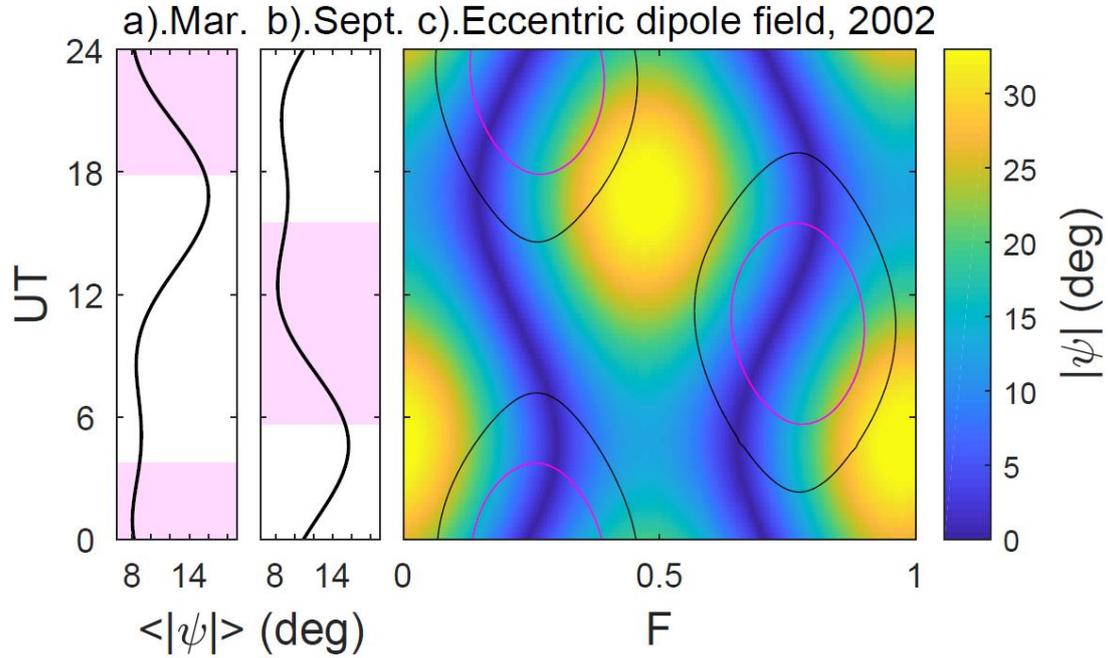

**Figure 8.** Comparison of the Russell-McPherron and equinoctial patterns, derived using an eccentric dipole geomagnetic field for the year 2002. In the *F-UT* plot in part c, the colour contours give the absolute value of the dipole tilt angle, $|\psi|$, superposed on which are contours showing the IMF orientation factor used by $P_\alpha$, namely $A_\theta = sin^4(\theta/2)$, where $\theta$ is the clock angle of the IMF in the GSM frame. These predictions are the average for an equal mix of $[B_Y]_{GSEQ} = -|B| < 0$ and $[B_Y]_{GSEQ} = +|B| > 0$ and contours are shown for $A_\theta$ of 0.28 (in black) and 0.31 (in mauve). Parts a and b show the mean values of $|\psi|$ over the range of *F* defined by the maximum extent in *F* of the mauve contour as a function of *UT*: Part a is for the March equinox and Part b for the September equinox. The area shaded peak is the *UT* extent of the peak defined by the mauve contour in c. The point of parts a and b is to show that the two equinoxes go through exactly the same sequence of variations in both $A_\theta$ and $\psi$ w *UT*, but with a phase difference of 12 hours.

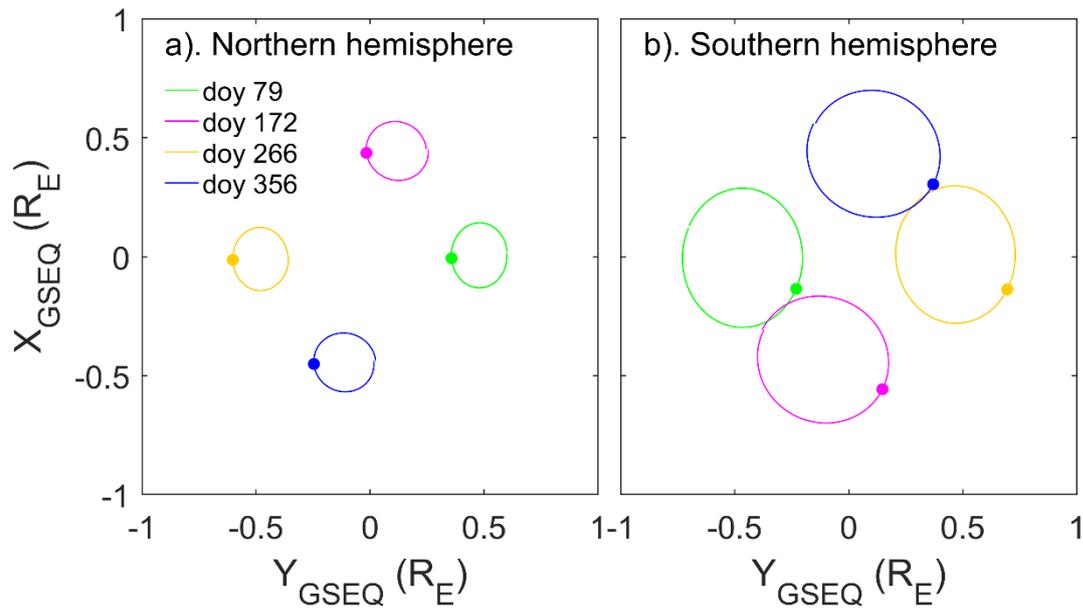

**Figure 9**. Locations of the geomagnetic eccentric dipole axial poles mapped into the *XY* plane of the GSEQ frame, viewed looking northward from the south of the solar equator (so that the Z GSEQ axis that makes up the right hand set points into the page) : (a) for the northern hemisphere pole, (b) for the southern hemisphere pole. The loci are shown for: (green) the March equinox (day of year, doy, 79, *F* = 0.21; (mauve) the June solstice (doy 172, *F* = 0.47); (orange) the September equinox (doy 266, *F* = 0.73); and (blue) the December solstice (doy 356, *F* = 0.98). The dots show the location at 12 *UT* in each case. Note that the $Z_{GSEQ}$ direction is into the plane of the diagram and so the *XY* plane is here viewed from the southern side and so dawn is the to the left and dusk to the right and noon to the top, (in the +*X* direction) and hence the motion of the poles with *UT* is clockwise for this viewpoint.

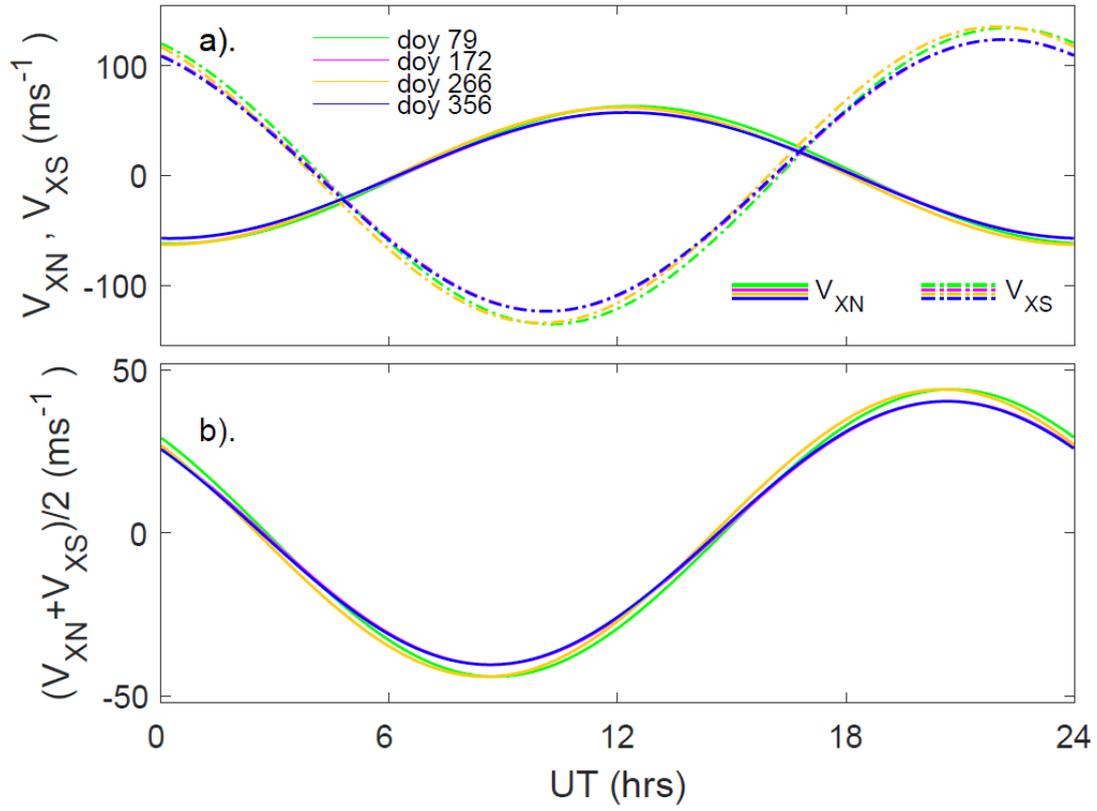

**Figure 10.** (a) Variations with *UT* for 2002 of the sunward velocity of the northern eccentric dipole axial pole, $V_{XN}$ (solid lines) and of the southern eccentric dipole axial pole northern , $V_{XS}$ (dot-dash lines). The variations are shown for the two solstices and the two equinoxes, using the same color scheme as Figure 9. (b) the average of the two, $V_{XNS}$.

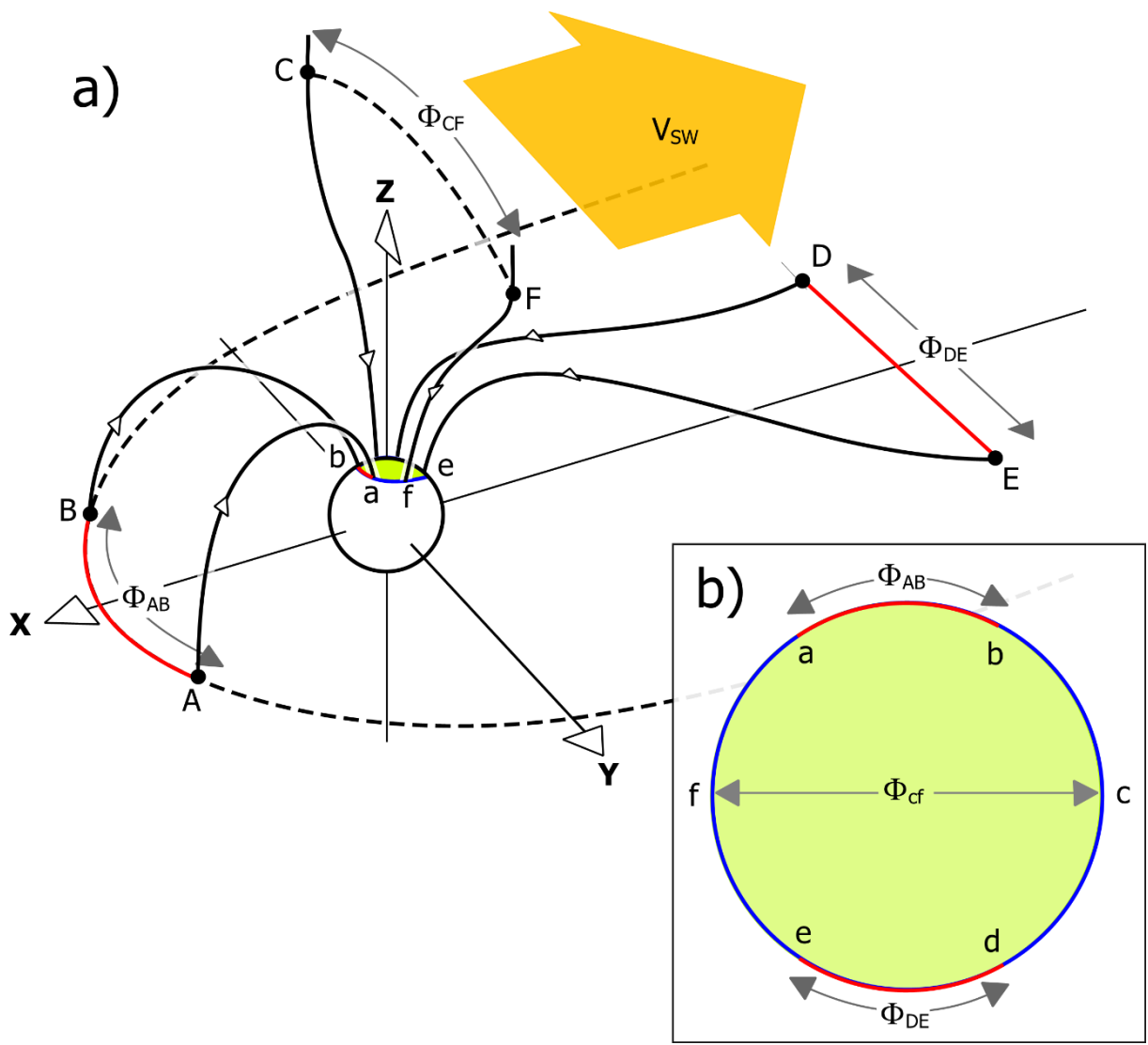

**Figure 11**. Schematic illustrating inductive decoupling of solar wind and ionospheric electric field and flows that is a key part of the Expanding-Contracting Polar Cap (ECPC) model of non-steady convection. In (a), the $X$, $Y$, and $Z$ axes of the Geocentric Solar Magnetospheric frame are shown. The points a, b, c, d, e and f are the ionospheric field line footpoints of the points on the magnetopause or cross tail current sheet A, B, C, D, E and F, respectively and all lie on the open-closed field line boundary, bounding the green area showing the open field line polar cap. AB is the dayside magnetopause reconnection X-line (across which the voltage $\Phi_{AB}$ is applied by the magnetic reconnection that opens field lines) and DE is the reconnection X-line in the cross tail current sheet (where the voltage $\Phi_{DE}$ is caused by reconnection that recloses open field lines). FC is the "Stern Gap" in interplanetary space, the ionospheric footprint of which is the polar cap of diameter, fc, and in which the open field lines are frozen-in to the solar wind flow, $V_{SW}$. (b) A view of the ionospheric polar cap (with noon at the top), with the green area again showing the open field line region, the blue lines showing "adiaroic" (non-reconnecting) segments of the open-closed boundary and red segments being "merging gaps" that map to the reconnection X-lines. (From *Lockwood and Morley*, 2004)

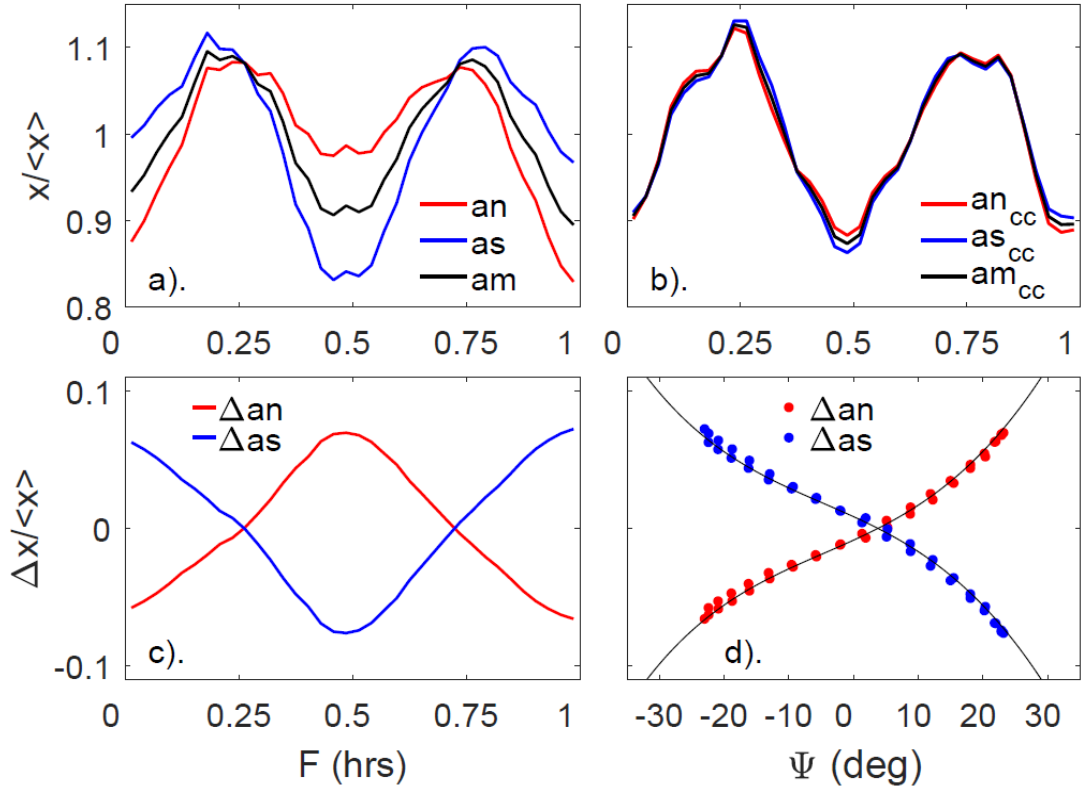

**Figure 12**. (a) The observed variations of the geomagnetic indices with fraction of year, $F$, shown as means in 36 equal-sized bins in $F$ as a fraction of their overall mean: (red) $an(F)/<an>_{all}$; (blue) $as(F)/<as>_{all}$; and (black) $am(F)/<am>_{all}$. (b) The variations after correction for conductivity effects: (red) $an_{cc}(F)/<an_{cc}>_{all}$; (blue) $as_{cc}(F)/<as_{cc}>_{all}$; and (black) $am_{cc}(F)/<am_{cc}>_{all}$. The deviations of $an$ and $as$ from $am$, (red) $\Delta an = an - am$, (blue) $\Delta as = as - am$, which for these variations with $F$ are taken to be due to conductivity effects alone. (d) The variations of (red points) $\Delta an$ and (blue points) $\Delta as$ as a function of the mean dipole tilt angle, $\psi$ for the same $F$-$UT$. The black lines in (d) are 4th-order polynomial fits to the points that are used to correct $an$ to $an_{cc}$ and $as$ to $as_{cc}$, using equations (7) and (10), respectively.

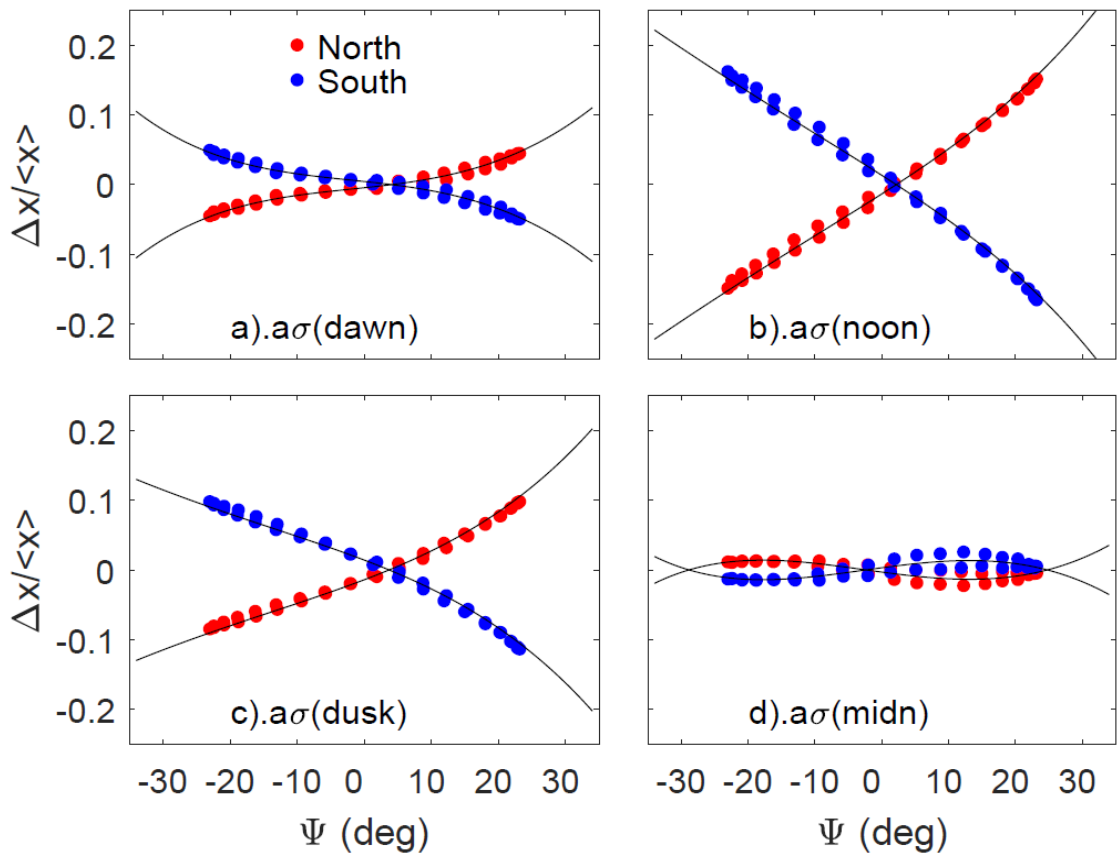

**Figure 13**. Plots corresponding to Figure 12d for the four $a\sigma$ indices: (a) $a\sigma(dawn)$; (b) $a\sigma(noon)$; (c) $a\sigma(dusk)$; and $a\sigma(midn)$. In each case the red/blue dots are for the northern/southern hemisphere component and the black lines are fourth-order polynomial fits.

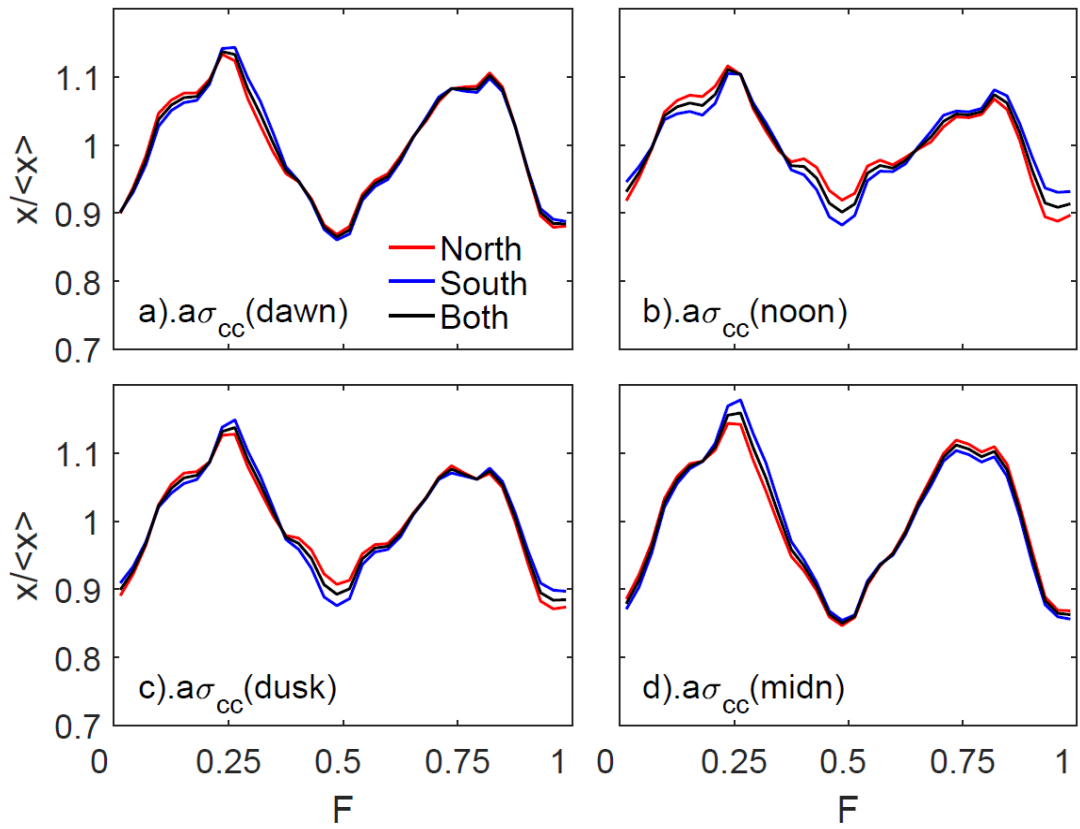

**Figure 14**. (a) The observed variations of the conductivity-corrected $a\sigma$ geomagnetic indices with fraction of year, $F$, shown as means in 36 equal-sized bins in $F$ as a fraction of their overall mean. In all panels, red lines are for the northern hemisphere index, blue for the south and black for the average of the two. (a) $a\sigma_{cc}(dawn)(F)/<a\sigma_{cc}(dawn)>_{all}$; (b) $a\sigma_{cc}(noon)(F)/<a\sigma_{cc}(noon)>_{all}$; (c) $a\sigma_{cc}(dusk)(F)/<a\sigma_{cc}(dusk)>_{all}$; and (d) $a\sigma_{cc}(midn)(F)/<a\sigma_{cc}(midn)>_{all}$.

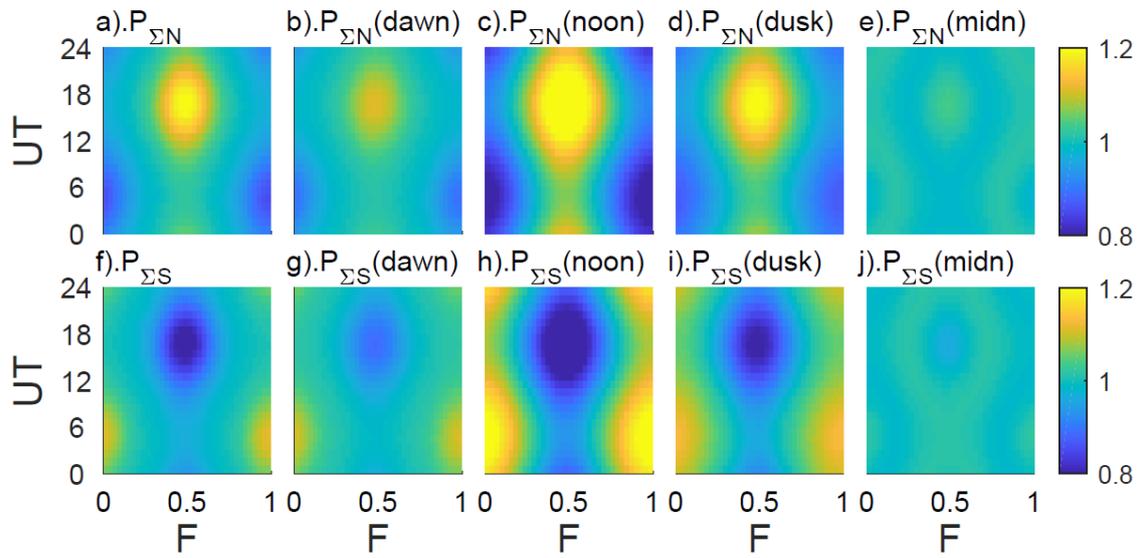

**Figure 15**. *F-UT* of conductivity factors for the hemispheric indices. The top row is for northern hemisphere indices, the bottom row for southern hemisphere indices. The columns from left to right are for: (a) and (f) for the hemispheric *an* and *as* indices, $P_{\Sigma N}$ and $P_{\Sigma S}$; (b) and (g) for the $a\sigma_N(dawn)$ and $a\sigma_S(dawn)$ indices, $P_{\Sigma N}(dawn)$ and $P_{\Sigma S}(dawn)$ ; (c) and (h) for the $a\sigma_N(noon)$ and $a\sigma_S(noon)$ indices, $P_{\Sigma N}(noon)$ and $P_{\Sigma S}(noon)$; (d) and (i) for the $a\sigma_N(dusk)$ and $a\sigma_S(dusk)$ indices, $P_{\Sigma N}(dusk)$ and $P_{\Sigma S}(dusk)$; and (e) and (j) for the $a\sigma_N(midn)$ and $a\sigma_S(midn)$ indices, $P_{\Sigma N}(midn)$ and $P_{\Sigma S}(midn)$.

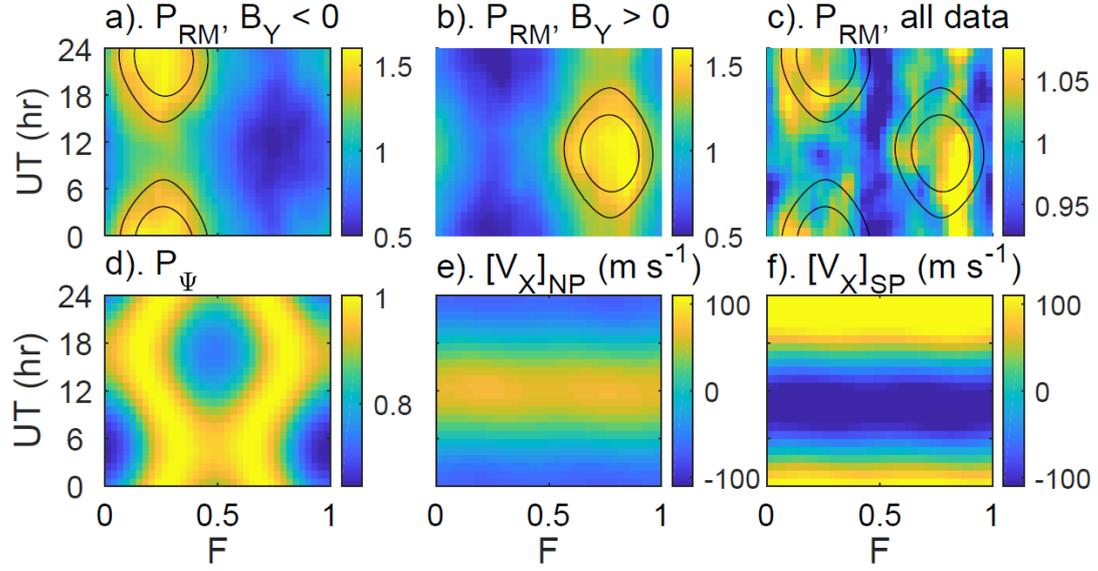

**Figure 16**. (a)-(c): *F-UT* plots of mean normalised power input into the magnetosphere, $P_\alpha/P_o$, average into 1-hour bins of *UT* and 36 equal-width bins of *F*. The data are 1-minute values averaged into 1-hour intervals using the criteria for handling data gaps that limits the errors they cause to ±5%, as defined by *Lockwood et al.* (2019b) and are sorted by the polarity of the *Y* component of the IMF over the prior hour: (a) is for $[B_Y]_{GSEQ} > 0$ and (b) is for $[B_Y]_{GSEQ} < 0$. The black contours are the locations of the peaks predicted for the Russell-McPherron effect, being the two contour lines for $A_\theta = 0.28$ and $A_\theta = 0.3$ plotted in Figure 8c. (d) is the *F-UT* plot for the $P_{\psi/\Delta B}(\psi)$ factor (see text) and (e) and (f) are the *F-UT* plots sunward velocity in the GSEQ frame of the axial geomagnetic poles at an altitude of 800 km in the northern and southern hemisphere, computed for each *F* and hourly *UT* in the same way as in Figure 10.

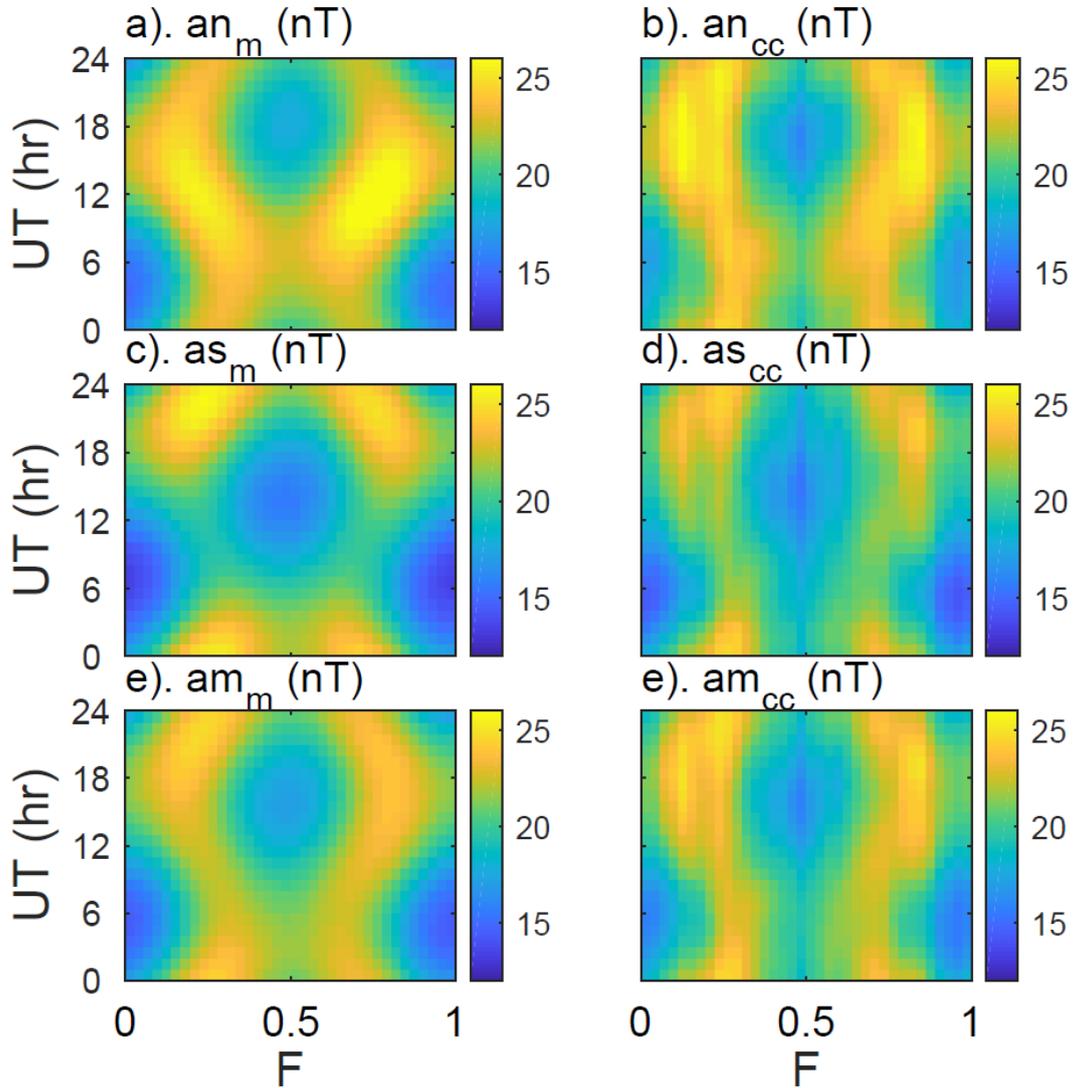

**Figure 17.** A comparison of *F-UT* patterns of (right column) conductivity-corrected observed indices and (left column) the corresponding modelled pattern, the top row is for the northern hemisphere *an* index, (a) $an_m$ and (b) $an_{cc} = an/P_{\Sigma N}$; the top middle row is for the southern hemisphere *as* index, (c) $as_m$ and (d) $as_{cc} = as/P_{\Sigma S}$; the bottom row is for the global hemisphere *am* index, (e) $am_m = (an_m + as_m)/2$ and (f) $am_{cc} = (an_{cc} + as_{cc})/2$. In all three cases the free fit parameter used is $c_{MP} = 0.27$, derived by minimising the r.m.s. fit residual for the *am* index case, $\{< (am_{cc} - am_m)^2 >\}^{1/2}$.

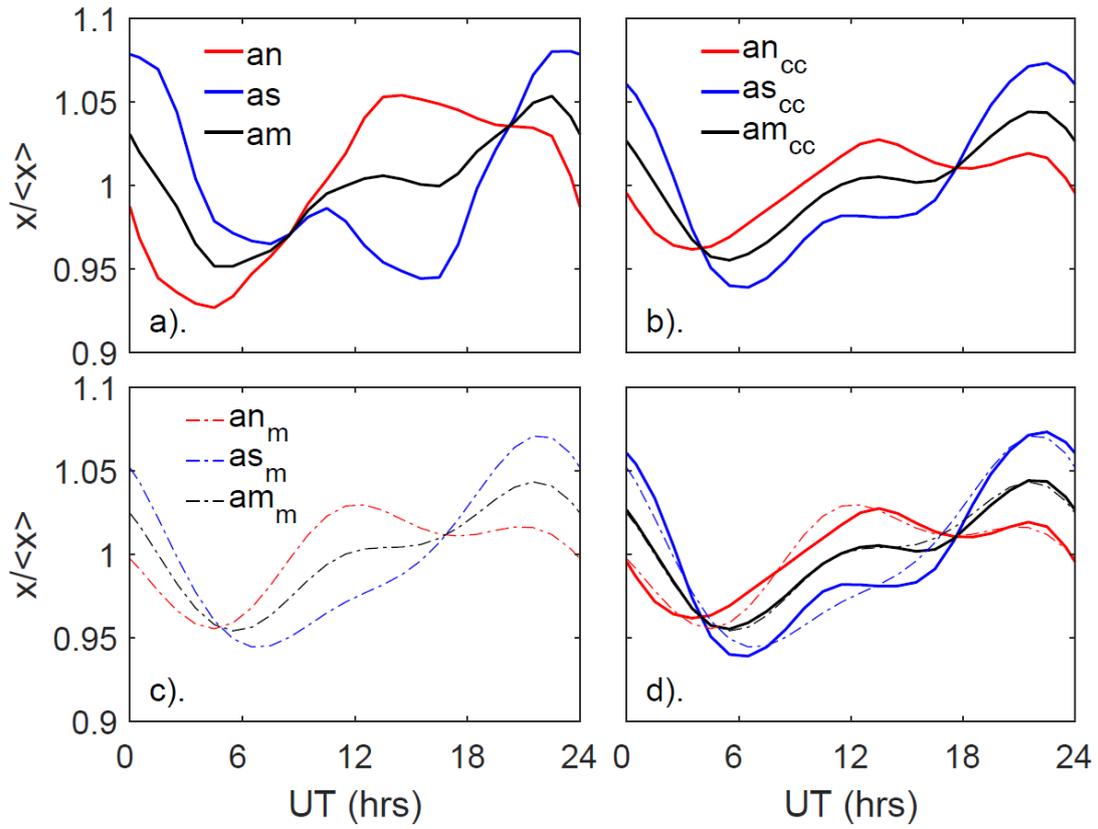

**Figure 18.** Observed, conductivity-corrected and modelled $UT$ variations for the (red) $an$, (blue) $as$, and (black) $am$ indices. (a) shows the values, $an$, $as$, and $am$ linearly interpolated from three hourly observations to 1-hour resolution and then averaged for the 24 1-hour $UT$ values for all data (for 1959-2019, inclusive). (b) shows the conductivity-corrected values $an_{cc} = an/P_{\Sigma N}$, $as_{cc} = as/P_{\Sigma S}$ and $am_{cc} = (an_{cc} + as_{cc})/2$. (c) shows the modelled values, $an_m$, $as_m$ and $am_m = (an_m + as_m)/2$. (d) compares the conductivity corrected values (solid lines) and the modelled values (dot-dash lines) on the same plot.

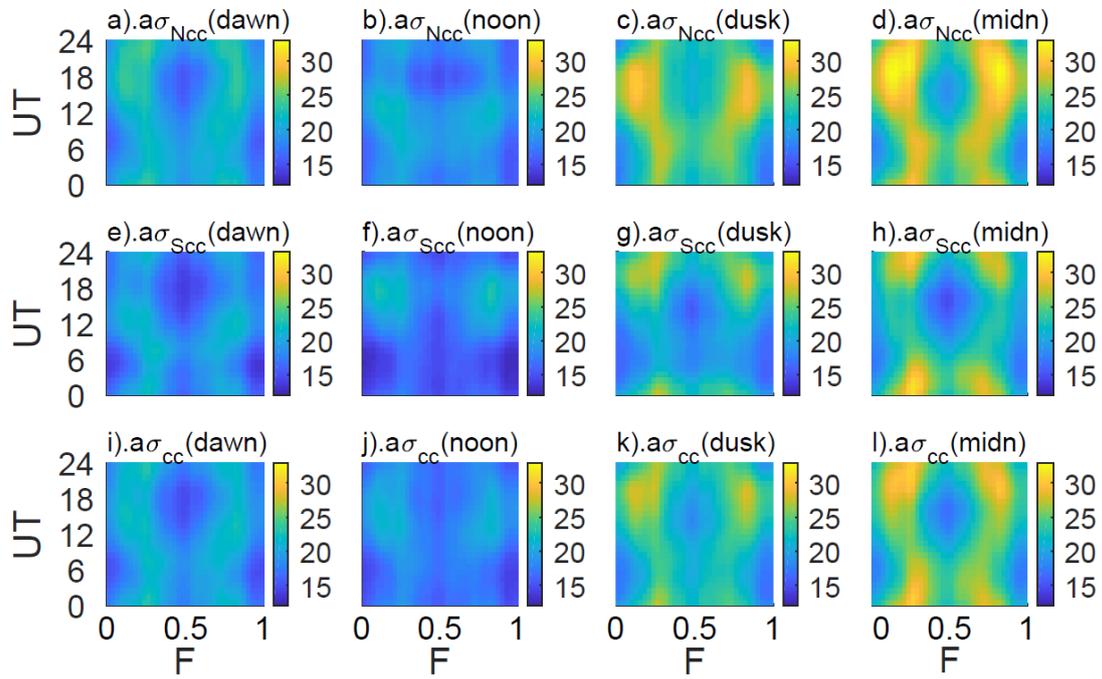

**Figure 19.** *F-UT* plots of the conductivity-corrected $a\sigma$ indices. The top row is for the northern hemisphere sub-indices, the middle row for the southern hemisphere sub-indices and the bottom row for the global indices. The columns are for the four 6-hour MLT sectors of the $a\sigma$ indices and from left to right are for dusk, noon, dawn and midnight.

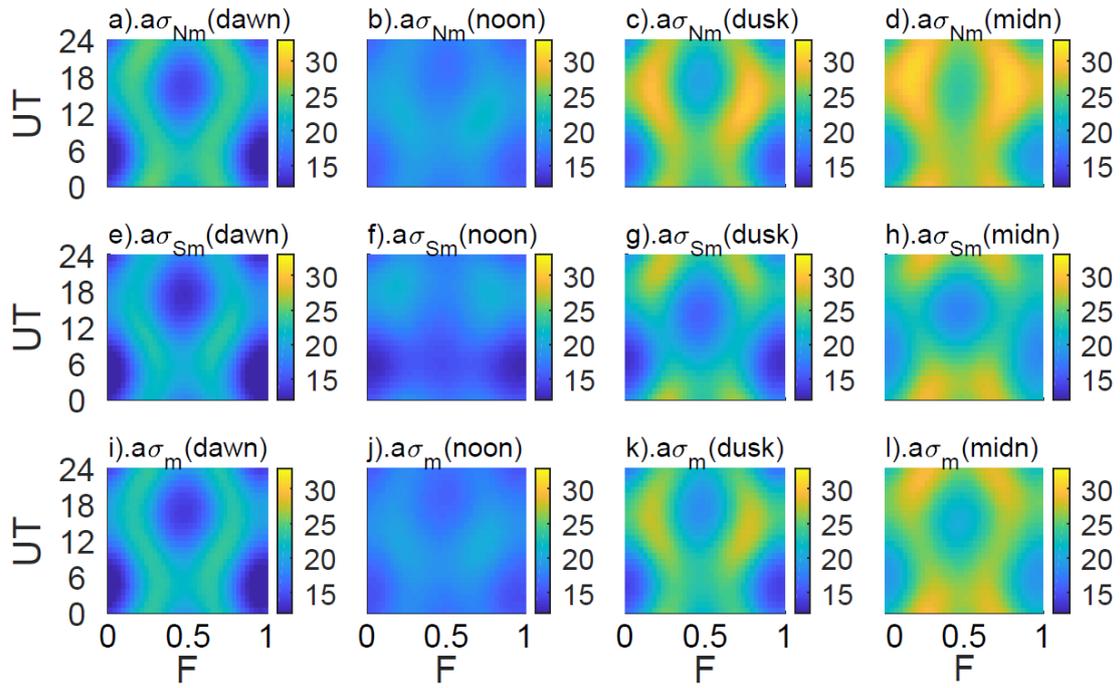

**Figure 20.** The same as Figure 19, for modelled values of the (conductivity-corrected) $a\sigma$ indices. The top row is for the northern hemisphere sub-indices, the middle row for the southern hemisphere sub-indices and the bottom row for the global indices. The columns are for the four 6-hour MLT sectors of the $a\sigma$ indices and from left to right are for dusk, noon, dawn and midnight.

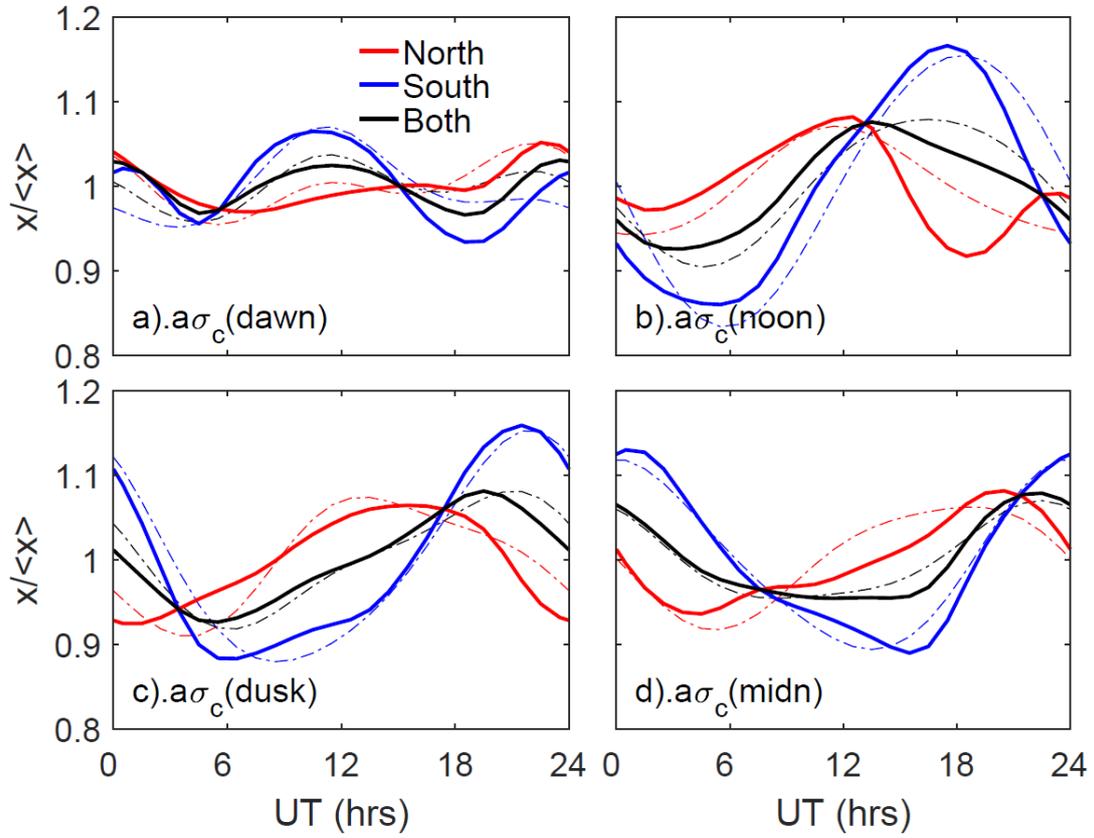

**Figure 21.** Comparison of conductivity-corrected and modelled *UT* variations for (red) northern hemisphere , (blue) southern hemisphere, and (black) global $a\sigma$ indices. Each panel compares the conductivity corrected values (solid lines) and the modelled values (dot-dash lines) on the same plot, as in Figure 18d. (a) for dawn, $a\sigma_{Ncc}(dawn)$, $a\sigma_{Scc}(dawn)$ and $a\sigma_{cc}(dawn)$ compared with $a\sigma_{Nm}(dawn)$, $a\sigma_{Sm}(dawn)$ and $a\sigma_m(dawn)$; (b) for noon, $a\sigma_{Ncc}(noon)$, $a\sigma_{Scc}(noon)$ and $a\sigma_{cc}(noon)$ compared with $a\sigma_{Nm}(noon)$, $a\sigma_{Sm}(noon)$ and $a\sigma_m(noon)$; (c) for dusk, $a\sigma_{Ncc}(dusk)$, $a\sigma_{Scc}(dusk)$ and $a\sigma_{cc}(dusk)$ compared with $a\sigma_{Nm}(dusk)$, $a\sigma_{Sm}(dusk)$ and $a\sigma_m(dusk)$; and (d) for midnight, $a\sigma_{Ncc}(midn)$, $a\sigma_{Scc}(midn)$ and $a\sigma_{cc}(midn)$ compared with $a\sigma_{Nm}(midn)$, $a\sigma_{Sm}(midn)$ and $a\sigma_m(midn)$.

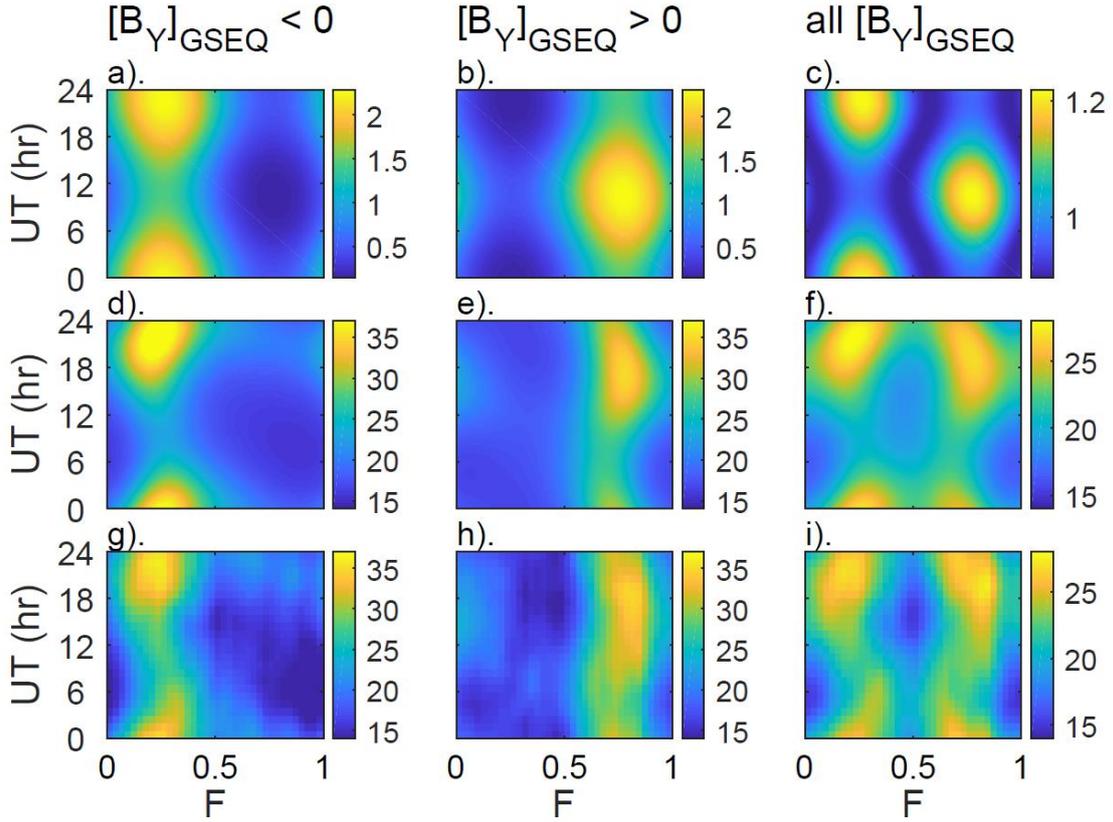

**Figure 22**. Simulations of $F$-$UT$ patterns sorted by the polarity of the average IMF $[B_Y]_{GSEQ}$ component during the prior hour. The top row shows the Russell-McPherron patterns $P_{RM}(F, UT)$ in normalised power input into the magnetosphere simulated using the eccentric dipole geomagnetic field model for 2002 with (a) $[B_Y]_{GSEQ} = +|B|$, (b) $[B_Y]_{GSEQ} = -|B|$, and (c) an equal mix of $[B_Y]_{GSEQ} = +|B|$ and $[B_Y]_{GSEQ} = -|B|$. The middle panels show the corresponding modelled patterns of $P_m(F, UT) = P_{RM}(F, UT).P_\psi(F, UT).P_{NS}(UT)$. The observed patterns in $a\sigma(midn)$ data shown in the bottom panels are for (g) $[B_Y]_{GSEQ} < 0$, (h) $[B_Y]_{GSEQ} > 0$ and (i) all data.

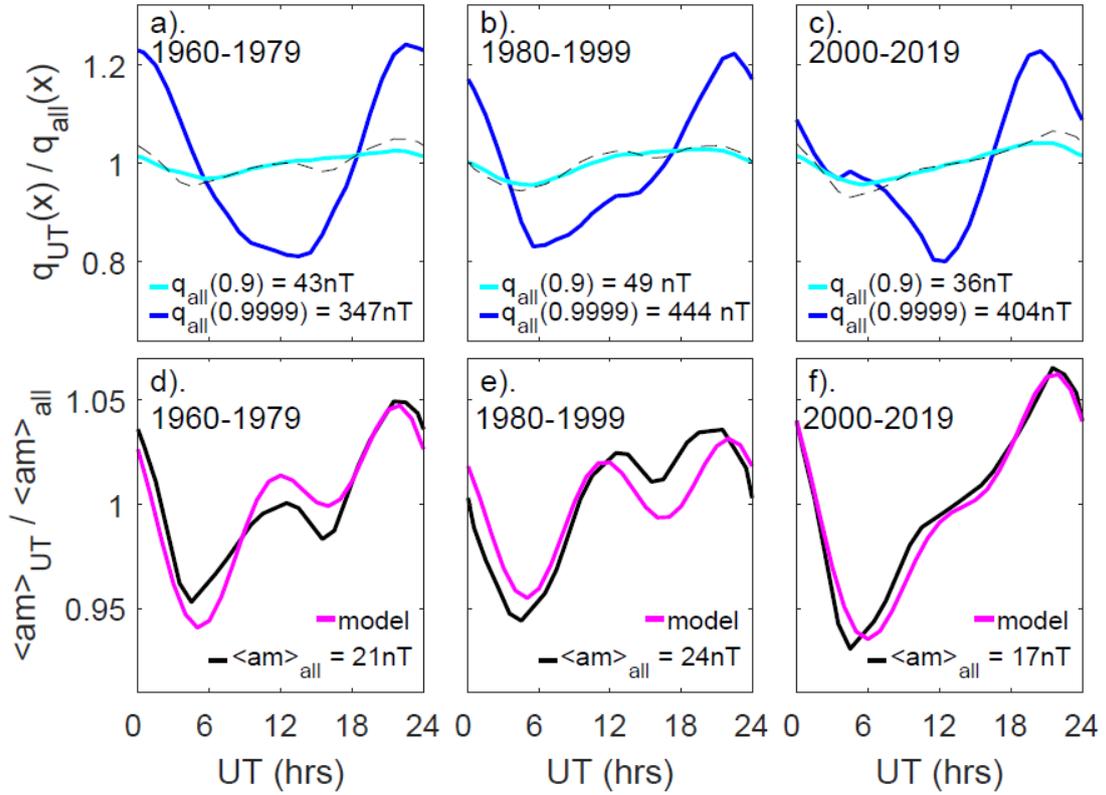

**Figure 23**. Universal time variations in *am* averaged over all *F* for: (a) and (d) 1960-1979; (b) and (e) 1980-1999; and (c) and (f) 2000-2019. The top panels show the *UT* variation in the 90% quantile of the distribution of *am* values (q(0.9), cyan lines) and the 99.99% quantile(q(0.9999), blue lines). The dashed lines show the variation for the mean *am*. Bottom panels show the variations for mean observed *am* (black lines) and modelled *am*, $am_m$ (mauve lines).

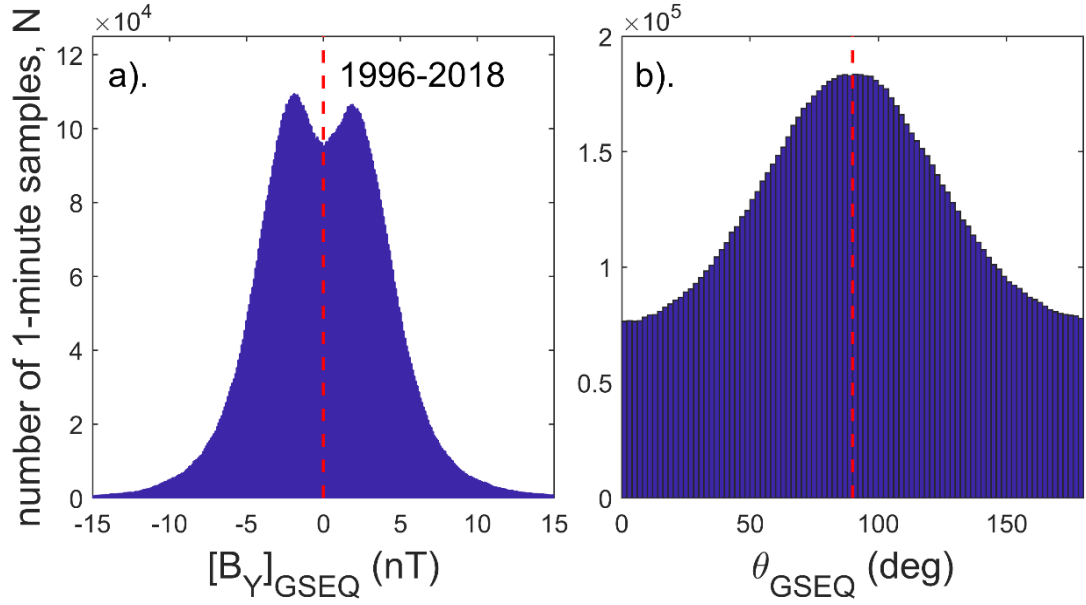

**Figure A1**. Distributions of IMF parameters for the full 22-year Hale cycle between 1996 and 2018: (a) the IMF $B_Y$ component, $[B_Y]_{GSEQ}$, the IMF clock angle in GSEQ, $\theta_{GSEQ} = tan^{-1}([B_Y]_{GSEQ}/[B_Z]_{GSEQ})$. In (a) the bins of $[B_Y]_{GSEQ}$ are 0.2nT wide and the vertical red dash line marks $[B_Y]_{GSEQ} = 0$; in (b) the bins of $\theta_{GSEQ}$ are 2° wide and the vertical dashed red line marks $\theta_{GSEQ}= 0$.

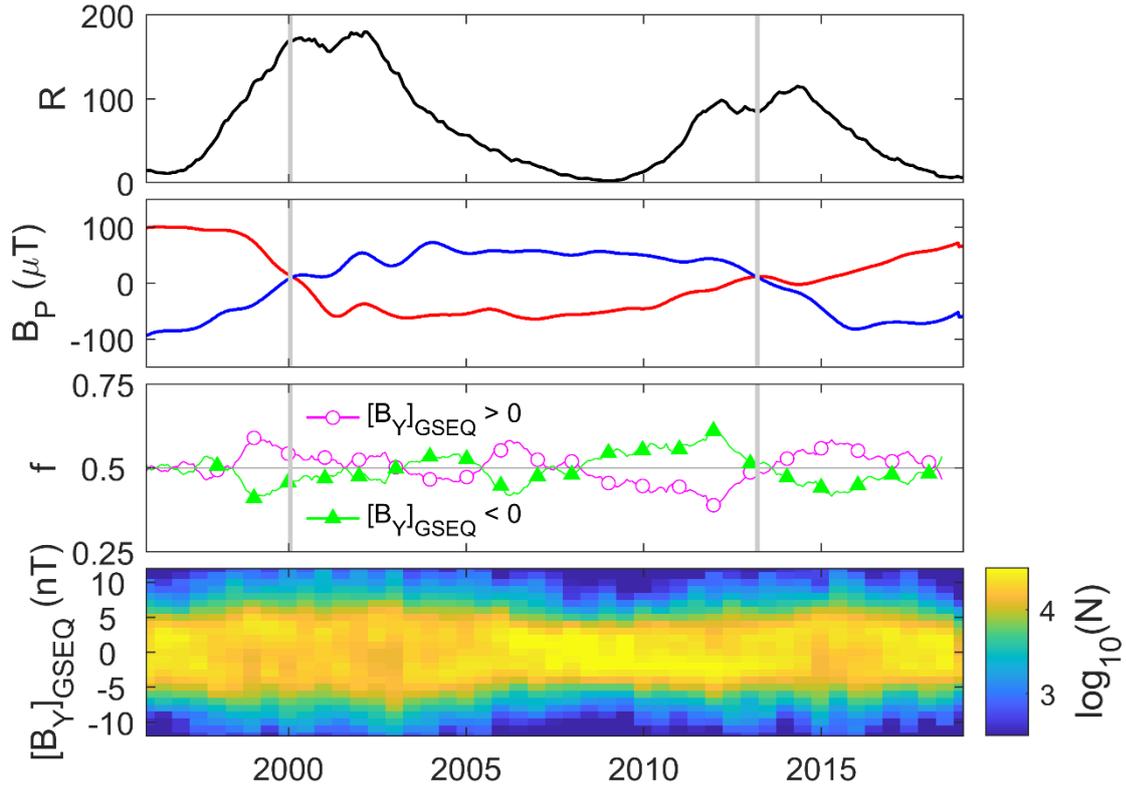

**Figure A2.** Variations with time of values over half-year intervals (centred on the equinoxes) for the 1996-2018 interval used in Figure A1. (a). The sunspot number, $R$. (b) The polar field strength from Wilcox Solar Observatory (WSO) magnetograms, $B_p$, where red/blue is for the north/south solar pole, respectively. (c) The fraction of time $f$ that the IMF has $[B_Y]_{GSEQ} > 0$ polarity (mauve line with open circles) and $[B_Y]_{GSEQ} < 0$ polarity (green line with solid circles). (d) The number of one-minute samples of $[B_Y]_{GSEQ}$ in bins 1nT wide and 6-month intervals centred on the equinoxes , $N$ to as a function of date and $[B_Y]_{GSEQ}$ (plotted on a logarithmic colour scale to reveal the tails of the distributions as well as the peaks). The vertical grey lines mark the polarity reversals of the solar polar fields

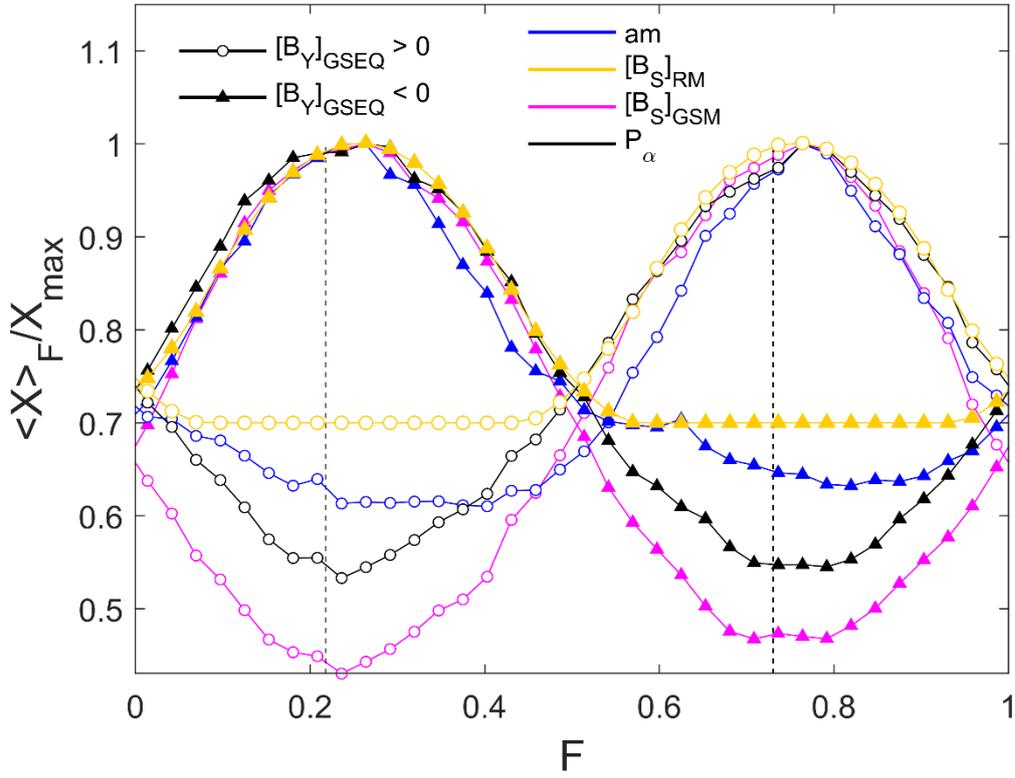

**Figure A3.** Variations with fraction of a calendar year, $F$, for the two polarities of the IMF $B_Y$ in the GSEQ frame. Normalised mean values, $< X >_F / X_{max}$, are shown in bins 1/36 yr wide, with open circles being for $[B_Y]_{GSEQ} > 0$ and solid triangles are for $[B_Y]_{GSEQ} < 0$. The data are for the interval 1996-2018. $< X >_F$ is the mean in each bin of $F$ and $X_{max}$ is the largest value of $< X >_F$ for a generic parameter $X$. The blue lines are for the $am$ index; the mauve lines are for the half-wave rectified southward field, $B_S$; the black lines are for the power input to the magnetosphere, $P_\alpha$; the orange line is the half-wave rectified southward field predicted for the R-M effect with IMF clock angle $[\theta]_{GSEQ} = 90°$, $[B_S]_{RM}$: $[B_S]_{RM} = b - [B_Z]_{RM}$ for $[B_z]_{RM} < 0$ and $[B_S]_{RM} = b$ for $[B_Z]_{RM} > 0$: and where $b$ is the best-fit baselevel value for northward IMF in GSM and $[B_Z]_{RM} = [B_Y]_{GSEQ} \sin(\beta_{GSEQ})$ is the northward field in GSM obtained by assuming $[\theta]_{GSEQ} = 90°$, $\beta_{GSEQ}$ being the rotation angle between the GSEQ and GSM frames.

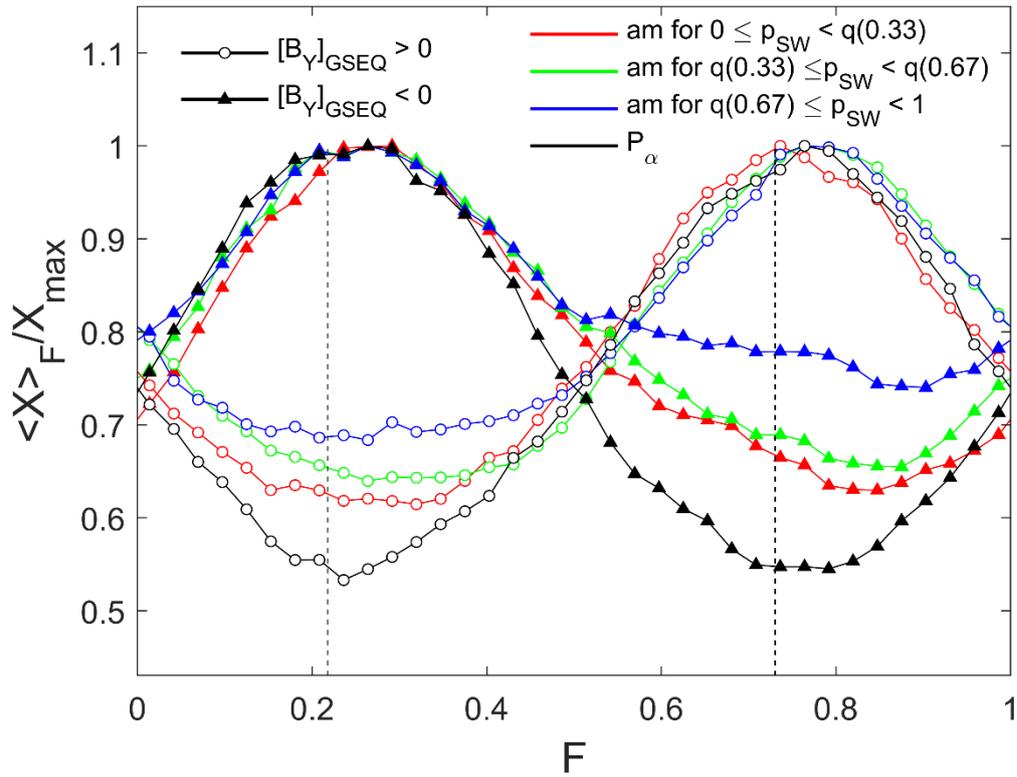

**Figure A4**. The same as Figure A3 for the *am* index for: (red lines) the lower tercile of the distribution of solar wind solar wind dynamic pressure, $p_{SW}$; (green lines) for the middle tercile of $p_{SW}$; and (blue lines) the upper tercile of $p_{SW}$. The black lines are for $P_\alpha$, as plotted in Figure A3, shown here for comparison.